\documentclass[preprint,12pt]{elsarticle}

\usepackage{amssymb, amsmath}
\usepackage[labelfont=bf, labelsep=space]{caption}
\usepackage[skip=0.5ex]{subcaption}
\usepackage{wrapfig}
\newcommand\scalemath[2]{\scalebox{#1}{\mbox{\ensuremath{\displaystyle #2}}}}

\journal{Journal of Theoretical Biology}

\begin{document}

\begin{frontmatter}



\title{Antigenic cooperation in Viral Populations: Transformation of Functions of Intra-Host Viral Variants\tnoteref{t1,t2}}


\author[inst1]{Leonid Bunimovich}
\ead{leonid.bunimovich@math.gatech.edu}

\affiliation[inst1]{organization={School of Mathematics, Georgia Institute of Technology},
            city={Atlanta},
            postcode={30332}, 
            state={Georgia},
            country={USA}}

\author[inst1,inst2]{Athulya Ram \corref{cor1}}\ead{athulya@gatech.edu}

\affiliation[inst2]{organization={Interdisciplinary Graduate Program in Quantitative Biosciences, Georgia Institute of Technology},
            city={Atlanta},
            postcode={30332}, 
            state={Georgia},
            country={USA}}

\author[inst3]{Pavel Skums}\ead{pskums@gsu.edu}

\affiliation[inst3]{organization={Department of Computer Science, Georgia State University},
            city={Atlanta},
            postcode={30302}, 
            state={Georgia},
            country={USA}}

\cortext[cor1]{Corresponding author}
\tnotetext[t1]{Declarations of interest: none}
\tnotetext[t2]{Acknowledgments: The authors acknowledge the support of Georgia Tech Interdisciplinary Graduate Program in Quantitative Biosciences. Funding: This work was partially supported by the NIH grant R01EB025022 and NSF grant CCF-2047828.}

\begin{abstract}
In this paper we study intra-host viral adaptation by antigenic cooperation - a mechanism of immune escape that serves as an alternative to the standard mechanism of escape by continuous genomic diversification and allows to explain a number of experimental observations associated with the establishment of chronic infections by highly mutable viruses.  Within this mechanism, the topology of a cross-immunoreactivity network forces intra-host viral variants to specialize for complementary roles and adapt to host's immune response as a quasi-social ecosystem. Here we study dynamical changes in immune adaptation caused by evolutionary and epidemiological events. First, we show that the emergence of a viral variant with altered antigenic features may result in a rapid re-arrangement of the viral ecosystem and a change in the roles played by existing viral variants. In particular, it may push the population under immune escape by genomic diversification towards the stable state of adaptation by antigenic cooperation. Next, we study the effect of a viral transmission between two chronically infected hosts, which results in merging of two intra-host viral populations in the state of stable immune-adapted equilibrium. In this case, we also describe how the newly formed viral population adapts to the host's environment by changing the functions of its members. The results are obtained analytically for minimal cross-immunoreactivity networks and numerically for larger populations.
\end{abstract}


\end{frontmatter}


\section{Introduction}\label{sec1}

RNA viruses such as HIV, Hepatitis C (HCV), Zika, Influenza A and SARS-CoV-2 are characterised by extremely high evolutionary rates \cite{drake1999mutation}. As a result each infected host or a community of infected individuals carries a heterogeneous population of genetically related viral variants \cite{domingo2012viral} that exist as an ecosystem, with the dominant selection pressure caused by hosts' immune systems \cite{Rhee07}. Until recently, the predominant model of viral evolution was the immune escape via continuous accumulation of genetic diversity \cite{nowak2000virus} often described as an ``arms race" between virus and hosts. However, several recent experimental discoveries are incompatible with the ``perpetual arms race” model. These discoveries include broad cross-immunoreactivity and antigenic convergence between intra-host viral variants \cite{campo2012hepatitis}, consistent increase in negative selection and decrease  of population heterogeneity over time \cite{ramachandran2011temporal,campo2014next,gismondi2013dynamic,lu2008hcv,illingworth2014identifying}, long-term persistence of viral variants \cite{ramachandran2011temporal,palmer2012insertion,palmer2014analysis} and complex fluctuations of frequencies of subpopulations over the course of infection \cite{ramachandran2011temporal,gismondi2013dynamic,palmer2014analysis,gray2012new,raghwani2016exceptional}. Given these observations, it is unlikely that the entire viral evolution is driven by a single evolutionary mechanism. It is rather a non-linear process defined by the recurring presentation of a succession of selection challenges specific to different stage of infection or epidemic spread \cite{baykal2020quantitative}. Each stage involves complex mechanisms which viruses share with other domains of life \cite{domingo2019social,baykal2020quantitative}.

One of the most intriguing phenomena of intra-host viral evolution is the transition between the immune escape under positive selection at early stage of infection and a conditionally stable state under the negative selection at the later state. Several previous modelling, genomic and experimental studies suggest that this transition can be caused by the development of specific cooperative interactions among viral variants \cite{skums2015antigenic,shirogane2013cooperation,domingo2019social,baykal2020quantitative} that allow viral populations to adapt to their environment as quasi-social systems \cite{domingo2019social}.  

The ODE model predicting and describing one possible scenario of such interactions has been proposed and analysed in our previous studies \cite{skums2015antigenic,bunimovich2019local,bunimovich2020local1,bunimovich2019specialization}. Cross-immunoreactivity network (CRN) plays a central role there. Although cross-immunoreactivity is essential for neutralization, its role is more complex and ambiguous. In particular, it does not
always act as a factor of pressure on virus,
but rather may serve as a factor facilitating virus survival through the mechanisms of  original antigenic sin,  heterologous immunity and antibody-dependent enhancement of viral infectivity \cite{francis1960doctrine,rehermann2005private,parsons2013benefits,meyer2008antibody}. The model assumes the presence of CRN with complex topology and takes into account a fundamental biochemical difference between antigenicity (capacity to bind antibodies) and immunogenicity (capacity to elicit antibodies) \cite{van2012basic,campo2012hepatitis,freitas1995role,mclean1997resource,tarlinton2006b,schwickert2007vivo,Pal12}. As a result, it describes a dynamic fitness landscape where viral variants determine fitnesses of other variants through their interactions in CRN. Antigenic cooperation and specialization of viral variants are naturally implied by the model as a way of mitigation of the immune pressure on certain antigenic variants at the expense of other variants. The state when the immune neutralization of particular variants is hampered was provisionally  called {\it local immunodeficiency} \cite{skums2015antigenic}.  The structure of CRN determines specific roles for each viral variant in host adaptation and local immunodefficiency emergence. Variants of high in-degrees play {\it altruistic} role and  improve fitness of adjacent variants at their own fitness cost by developing a polyspecific antibody response that interferes with development of specific immune responses against other variants immunoreactive with these antibodies. The latter variants are {\it selfish} because they gain fitness at the expense of in-hub variants. Thus, the model describes a cooperation between neighbors in CRN which correspond to altruism through kin selection \cite{hamilton1964genetical}.  This mechanism allows to explain a number of empirical observations. It is also stable and robust under various realistic conditions \cite{bunimovich2019local}.  

In contrast,  alternative models \cite{nowak2000virus,wodarz2003hepatitis,iwasa2004some}  suggest that immune escape is associated just with increase in genomic heterogeneity, which  is often inconsistent with the experimental observations. Notably, the antigenic cooperation model achieves its predictive power by using fewer variables than most of the previously proposed models \citep{wodarz2003,nowak_etal1990, nowak_etal1991, nowak_may1991}. The reasons for that is that are (a) high non-linearity of the model that allows to capture non-linear evolutionary effects; (b) more delicate exploration of the effects of cross-immunoreactivity via introduction of CRN with a complex topology as a model parameter, in contrast to mean-field approximation of immune responses utilized by many existing models. 

Antigenic cooperation model has been rigorously studied in several prior papers. The original paper \cite{skums2015antigenic}, besides introducing the model, described the emergence of antigenic cooperation and local immunodeficiency as its inherent properties using both numerical simulations and analytical exploration of its equilibrium solutions. The paper \citep{BS1} demonstrated that solutions implying local immunodeficiency can be stable
and robust under various realistic conditions for several specific types of cross-immunoreactivity networks. Another paper \citep{BS2} studied the role of altruistic viral variants in intra-host adaptation. It demonstrated that without altruistic variants the viral population could maintain only marginally stable state of a local immunodeficiency and a relatively small size. 

However, viral populations and, consequently, cross-immunoreactivity networks are not static and are subject to dynamical changes caused by emergence or introduction of viral variants with altered phenotypes  This fact raises a fundamental question: 
whether or how changes in CRNs lead to evolutionary transitions and, in particular, what are the effects of such changes on the functions of specific viral variants and on the immune escape of the entire population?

This question is in the focus of the present paper. We study dynamical changes in immune adaptation caused by two types
evolutionary and epidemiological events: (a) emergence of a new viral variant with altered antigenic phenotype and (b) a viral transmission between two chronically infected hosts, which results in merging of two intra-host viral populations in the state of stable immune-adapted equilibrium. Both phenomena are typical for evolution of the intra-host viral populations and important for understanding the laws of their evolution. 

We analyse these processes statically, assuming that emergence of new antigenic variants occurs in a given state of a virus-host system, and analysing what will be a new stable state of the system. This new stable state will be (formally) achieved in an infinite time. Then we study this process of transition from an ``old" state to the new one dynamically by following the previous evolution of the initial network and then its (future) dynamics after the emergence of new variants. 

It turned out that such events, may result in a rapid re-arrangement of the viral ecosystem and a change of the roles played by viral variants. In addition, it is rigorously demonstrated that emergent antigenic variants  may successfully co-exist with present persistent variants and become persistent itself while keeping the state of stable local immunodeficiency in the CRN. Another, less expected, and potentially more important finding, is that emergence of new variants may push the population under immune escape by genomic diversification towards the stable state of adaptation by antigenic cooperation. These findings emphasise how phenotypic features of particular viral genomic variants are formed by both  their antibody and ``quasi-social" environments rather then pre-defined by their genomes. They also highlight challenges in effective vaccine design by demonstrating how the evolutionary trajectories of intra-host viral populations subjected to the introduction of new antigenic variants are affected by the state of pre-existing populations.

The paper is organized as follows. In the next section we present a basic model of intra-host viral evolution in presence of a complex cross-immunoreactivity network. Section 3 is dealing with the transformations which result by emergence of a new viral variants in the population under a stable state of local immunodeficiency. In Section 4 we analyse the process of the union of two CRNs each having a stable state of LI.
All technical computations are presented in the Appendix. 

\section{Model of evolution of intra-host viral population organized into heterogeneous cross-immunoreactivity network}\label{sec:model}

In this section we describe the mathematical model of the viral population organized into heterogeneous cross-immunoreactivity network. The model was introduced in \cite{skums2015antigenic} and applied to Hepatitis C virus, but is applicable to any highly mutable pathogen with broad spectrum of cross-immunoreactivity. We consider a population of $n$ viral antigenic variants $x_i$ inducing $n$ immune responses $r_i$ in the form of antibodies and memory B-cells. We assume that viral variants form a cross-immunoreactivity network.
This network can be represented as a weighted directed graph
$G_{CRN} = (V, E)$, with vertices corresponding to viral variants and a pair of vertices $u$ and $v$ being connected by an arc, if an antigen $u$ interacts with $v$-specific antibodies elicited by $v$ as an immunogen. We incorporate the asymmetry between immune activation and neutralization into the model by considering two weight functions for the edges of $G_{CRN}$. These functions are described by immune neutralization and immune stimulation matrices $U = (u_{i,j})_{i,j=1}^n$ and
$V = (v_{i,j})_{i,j=1}^n$, where: $0 \leq u_{i,j},v_{i,j} \leq 1$; 
$u_{j,i}$ is a coefficient representing the binding affinity of
antibodies $r_j$ with $i$th variant; and $v_{i,j}$ is a coefficient reflecting strength of stimulation of antibodies to $r_j$ by $i$th variant. The immune response $r_i$ against the variant $x_i$ is
neutralizing; i.e., $u_{ii} = v_{ii} = 1$.

The resulting viral and antibody population dynamics is
described by the following system of ordinary differential equations:

\begin{equation}\label{population}
\begin{split}
\dot x_i=f_ix_i-px_i\sum_{j=1}^nu_{ji}r_j,\quad i=1,\dots,n,\\
\dot r_i=c\sum_{j=1}^nx_j\frac{v_{ji}r_i}{\sum_{k=1}^nv_{jk}r_k}-br_i,\quad i=1,\dots,n.
\end{split}
\end{equation}

In this model
the viral variant $x_i$ replicates at the rate $f_i$ and is eliminated by the immune responses $r_j$ at the rates $pu_{ji}r_j$,where $p$ is a constant. It is known that $x_j$ preferentially stimulates pre-existing immune responses capable of binding to $x_j$ with a relatively high affinity \cite{nara2010can} (or, in other words, immune responses {\it compete} for stimulation). It provides a rapid secondary immune response to re-infections with the same pathogen, but also results in the original antigenic sin,  repertoire freeze and heterologous immunity \cite{francis1960doctrine,kim2009original,midgley2011depth,parsons2013benefits,rehermann2005private}. This phenomenon is incorporated into the model by assuming that immune responses $r_i$ are stimulated by the $j$-th variant at the rates $cg_{ji}x_j$, where $g_{ji}=\frac{v_{ji}r_i}{\sum_{k=1}^nv_{jk}r_k}$ represents probability of stimulation of the immune response $r_i$ by the variant $x_j$, and $c$ is a constant. Without stimulation, immune responses $r_i$ decay at the rate $b$.

Similarly to \citep{skums2015algorithms,BS1}, here we are mostly interested in the effects of the CRN structure (topology) on the population dynamics. Thus
we consider the situation where the immune stimulation and neutralization coefficients are equal to constants $\alpha$ and $\beta$, respectively. In this case, we have
$$U=\text{Id}+\beta A^T,V=\text{Id}+\alpha A,$$
where $A$ is the adjacency matrix of the graph $G_{CRN}$. In numerical simulations, we assume that $0<\beta=\alpha^k$, where $k$ is the number of epitopes that should be binded for neutralization.  

Note that in the absence of cross-immunoreactivity, the system (\ref{population}) reduces to the model described in \citep{nowak_may2000}. In that case, equilibrium sizes of populations of viral variants and immune responses are 

\begin{equation}\label{eq:mainmodel}
    x_i^{\circ} = \frac{bf_i}{cp}, r_i^{\circ} = \frac{f_i}{p}, 
\end{equation}

One of the most interesting properties of the system (\ref{population}) is the emergence of so-called state of {\it local immunodefficiency}. It is defined as an equilibrium solution $(\mathbf{x^*},\mathbf{r^*})$ such that every viral variant $i$ falls into one the following 3 categories:

\begin{itemize}
\item[1)] $x_i^* > 0$ and $r_i^* \leq r_i^{\circ}$ ({\it persistent variants});

\item[2)] $x_i^* = 0$ and $r_i^* > 0$ ({\it altruistic variants});

\item[3)] $x_i^* = r_i^* = 0$ ({\it transient variants}).
\end{itemize}

Transient variants are being eliminated by the host's immune system as they emerge, and thus are subject to the standard immune escape by continuous diversification mechanism. The relations between persistent and altruistic variants are more interesting, as they describe a different mechanism of immune escape by antigenic cooperation. Under this mechanism, persistent variants survive without eliciting any specific immune responses (the state of "local immunodefficiency" with respect to these variants, whereat the immune system effectively "does not see" them). This is achieved via the agency of altruistic variants that does not survive but support the continuous existence of persistent variants. The roles of viral variants in this scheme are defined by their position in the CRN, with altruistic variants usually (but not always) being network hubs, and persistent variants being adjacent to them. Qualitatively, the mechanism can be described as follows. Under the model (\ref{population}), if the viral variant $x_i$ is adjacent to an altruistic variant $x_j$, then the immune response $r_i$  competes for activation with the immune response $r_j$. Since the latter response is broadly cross-immunoreactive and being stimulated by many variants, even after elimination of $x_i$ ($x_i = 0$) , it is preserved and readily outcompetes the former response, thus preventing it from development ($r_j = 0$). At the same time, $r_j$-antibodies  may lack sufficiently high affinity to neutralise $x_i$, which leads to its persistence ($x_j > 0$). One can consider these interactions as a form of cooperation between altruistic and persistent variants, where the former lose their fitness by significantly contributing to the fitness of the latter.

The state of local immunodefficiency, when exist, is usually stable and robust, as was confirmed both analytically and numerically \cite{skums2015antigenic,BS1,BS2}.

\section{Emergence of a new viral variant}

This section deals with the situation when a new variant is added to a cross-immunoreactivity network.  We found that as a result, roles of viral variants may change in different ways.

\subsection{Adding a new viral variant to a minimal branch-cycle network}\label{sec:adding}
\begin{figure}[ht]
    \centering
    \includegraphics[scale=0.3]{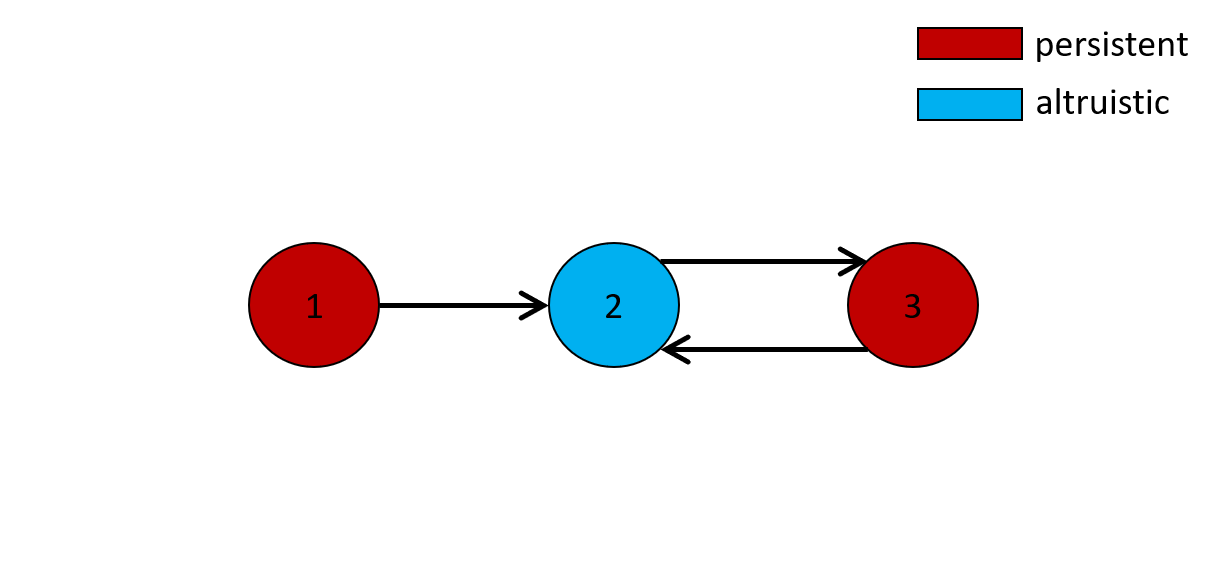}
    \caption{Stable configuration of the branch-cycle network. Node categories are highlighted in different colors}
    \label{fig:branch_cycle}
\end{figure}

A branch-cycle network is one of just the two smallest CR networks \citep{BS1,BS2} which can exhibit the property of a stable and robust local immunodeficiency.  The network and the roles of viral variants in the corresponding solution (that is derived and described in  \citep{BS1,BS2}) are depicted on Fig. \ref{fig:branch_cycle}. We analyzed all possible additions of a new node to this network and the resulting equilibrium solutions. 

\begin{figure}[ht]
     \centering
     \begin{subfigure}[b]{0.45\textwidth}
         \centering
         \includegraphics[width=\textwidth]{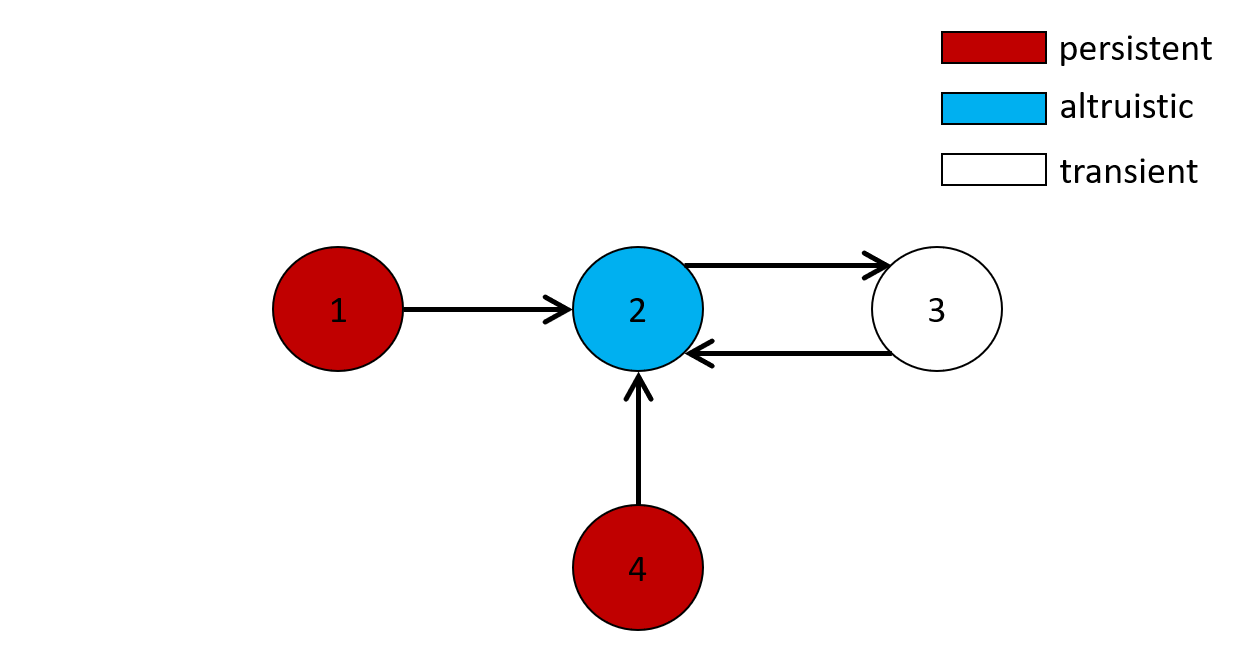}
         \caption{}
         \label{fig:per4_CRN1_fp2}
     \end{subfigure}
     \hfill
     \begin{subfigure}[b]{0.45\textwidth}
         \centering
         \includegraphics[width=\textwidth]{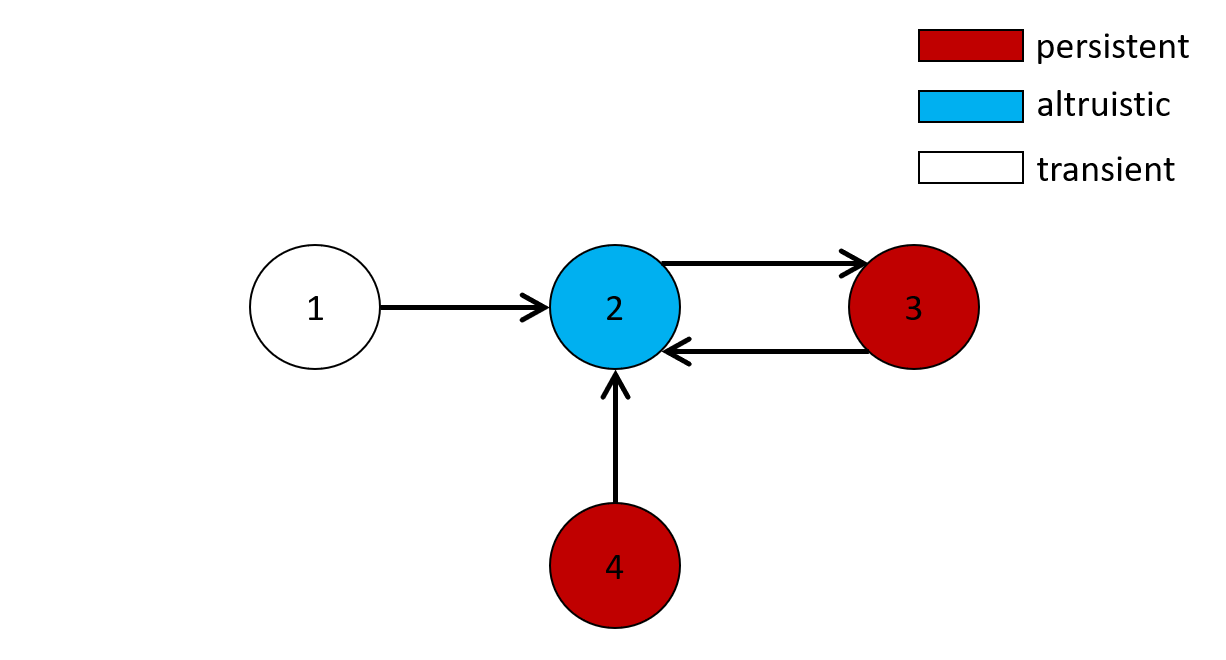}
         \caption{}
         \label{fig:per4_CRN1_fp3}
     \end{subfigure}
        \caption{Stable states where the emerging viral variant 4 becomes persistent in a branch-cycle network}
        \label{fig:per4_CRN1}
\end{figure}

The most notable finding is the existence of solutions where introduction of a new node changes functions of preexisting variants. All such stable solutions are shown on figure (\ref{fig:per4_CRN1}); in all cases the change occur when a new variant (node 4) is linked to the altruistic variant of the previous configuration. Other cases (where the newly emerged viruses are connected to a pre-existing persistent virus) is detailed in appendix \ref{appendix3}.

The fixed point shown in figure \ref{fig:per4_CRN1_fp2} corresponds to the following solution:  

\begin{flalign*}
    x_1^* &= \frac{b(\beta f_1 + (\alpha - \beta)f_4)}{\beta c p}, &x_2^* &= 0, &x_3^* &= 0, &x_4^* &= \frac{b f_4(1 - \alpha)}{\beta c p}\\
    r_1^* &= \frac{f_1 - f_4}{p}, &r_2^* &= \frac{f_4}{\beta p}, &r_3^* &= 0, &r_4^* &= 0
\end{flalign*}

This fixed point is stable under the conditions $\alpha > \frac{1}{2}, f_1 > f_4, f_4 > f_3,$ and $f_4 > \beta f_2$ (see Supplement)

The second fixed point shown in figure \ref{fig:per4_CRN1_fp3} corresponds to 
\begin{flalign*}
    x_1 &= 0, &x_2 &= 0, &x_3 &= \frac{b(\beta f_3 + (\alpha - \beta)f_4)}{\beta c p}, &x_4 &= \frac{b f_4(1 - \alpha)}{\beta c p}\\
    r_1 &= 0, &r_2 &= \frac{f_4}{\beta p}, &r_3 &= \frac{f_3 - f_4}{p},&r_4 &= 0
\end{flalign*}

This fixed point is stable if $\alpha > \frac{1}{2}, f_4 > f_1, f_3 > f_4,$ and $f_4 > \beta f_2$ (see Supplement)

Changes described by these two solutions are structurally similar. In both cases newly added variant becomes persistent, while the previously persistent variant is eliminated by the immune system and the altruistic variant retains its role. A necessary condition for stability of such qualitative changes of viruses functions is that the replication rate of a emergent variant is greater than that of the preexisting persistent variant. 

 Notably, in both cases the change occur in the variant not adjacent to the newly added variant. It demonstrates how network-mediated interactions between viral variants  propagate along the cross-immunoreactivity networks and thus go beyond direct interactions described in Section \ref{sec:model}. In this particular case, we observe a natural selection acting on potentially persistent variants supported by the same altruistic variant, with the variant of the lower fitness being eliminated and replaced by the newly emerged variant. 
 


The fact that the newly emerging variant is cross-immunoreactive as antigen with the immune response against pre-existing altruistic variant is essential.  Indeed, when the variant 4 is cross-immunoreactive with variants 1 or 3, then it either becomes transient while the roles of pre-existing variants are unchanged, or the dynamics of the CR network becomes unstable, i.e. it does not have a stable and robust state of local immunodeficiency.

\subsection{Adding a new viral variant to a minimal symmetric network}\label{sec:adding1}
\begin{figure}[ht]
    \centering
    \includegraphics[width=0.45\textwidth]{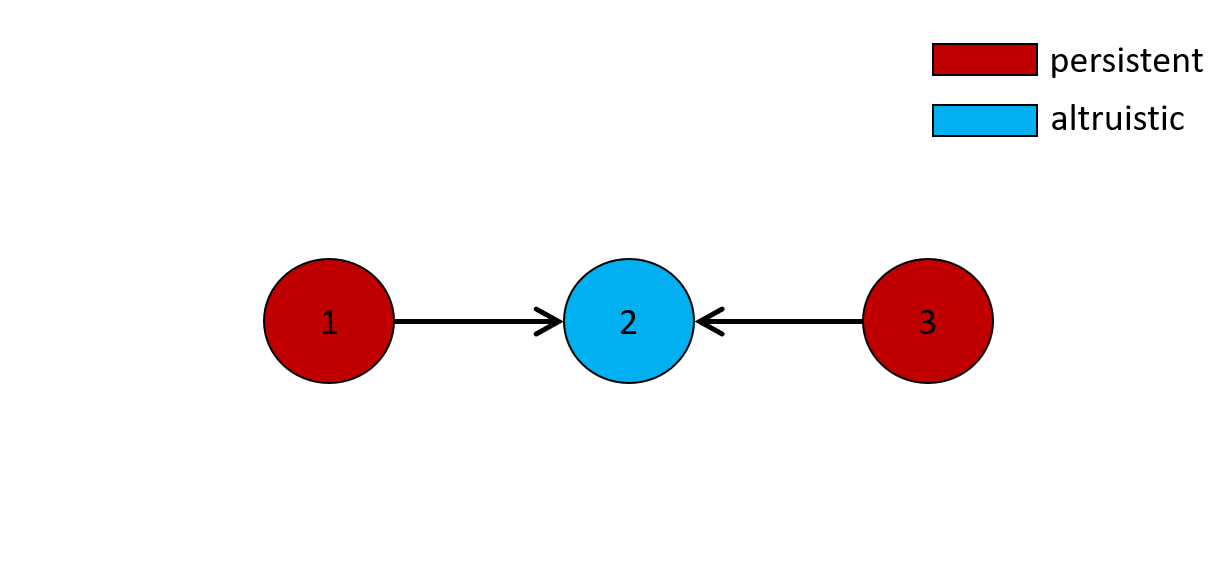}
    \caption{Symmetric minimal network}
    \label{fig:symmetricCRN}
\end{figure}

A symmetric network \citep{BS2} is another instance of the two smallest CR networks that can exhibit stable state of local immunodeficiency (Fig. \ref{fig:symmetricCRN}).

It was proven in \citep{BS2} that a stable state of LI exists in this network if $f_3 > f_1$, with the fixed point corresponding to
\begin{flalign*}
    x_1 &= \frac{bf_1}{cp \beta}(1 - \alpha), &x_2 &= 0, &x_3 &= \frac{b}{cp \beta}(\alpha f_1 + \beta (f_3 - f_1))\\
    r_1 &= 0, &r_2 &= \frac{f_1}{p \beta}, &r_3 &= \frac{f_3 - f_1}{p}
\end{flalign*}
As this network is symmetric, there is a similar fixed point with the switched solutions for variants 1 and 3; that solution is stable under the condition $f_1 > f_3$.

\begin{figure}[ht]
     \centering
     \begin{subfigure}[b]{0.45\textwidth}
         \centering
         \includegraphics[width=\textwidth]{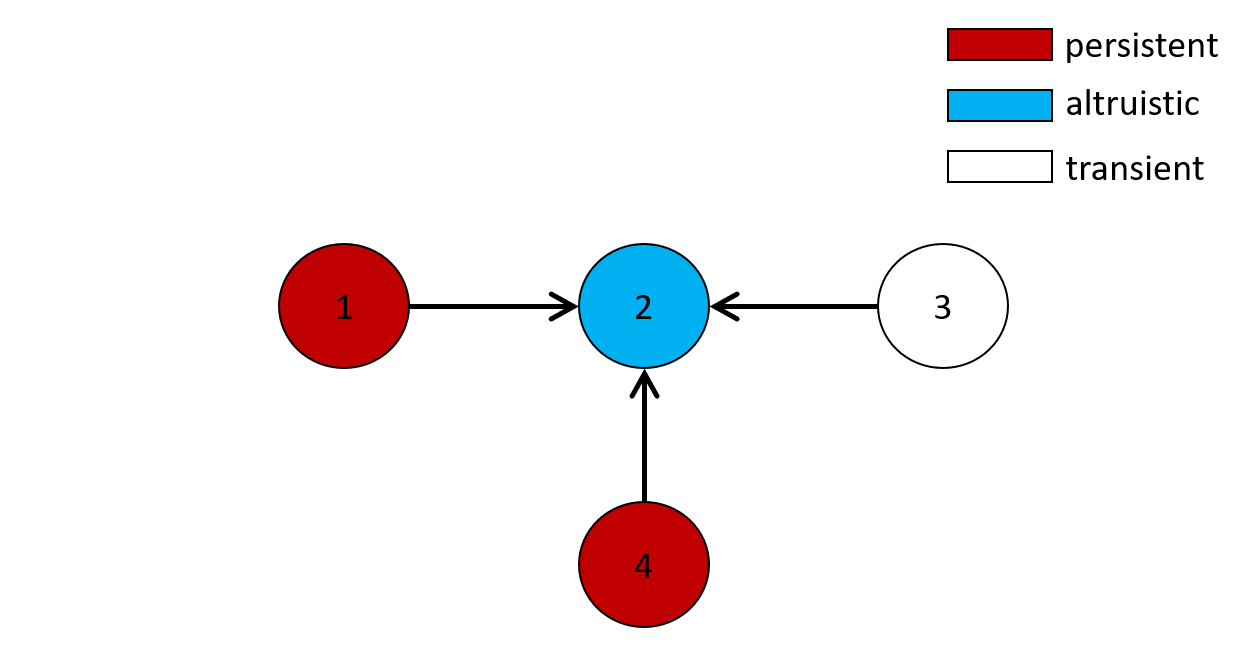}
         \caption{}
         \label{fig:per4_CRN2_fp2}
     \end{subfigure}
     \hfill
     \begin{subfigure}[b]{0.45\textwidth}
         \centering
         \includegraphics[width=\textwidth]{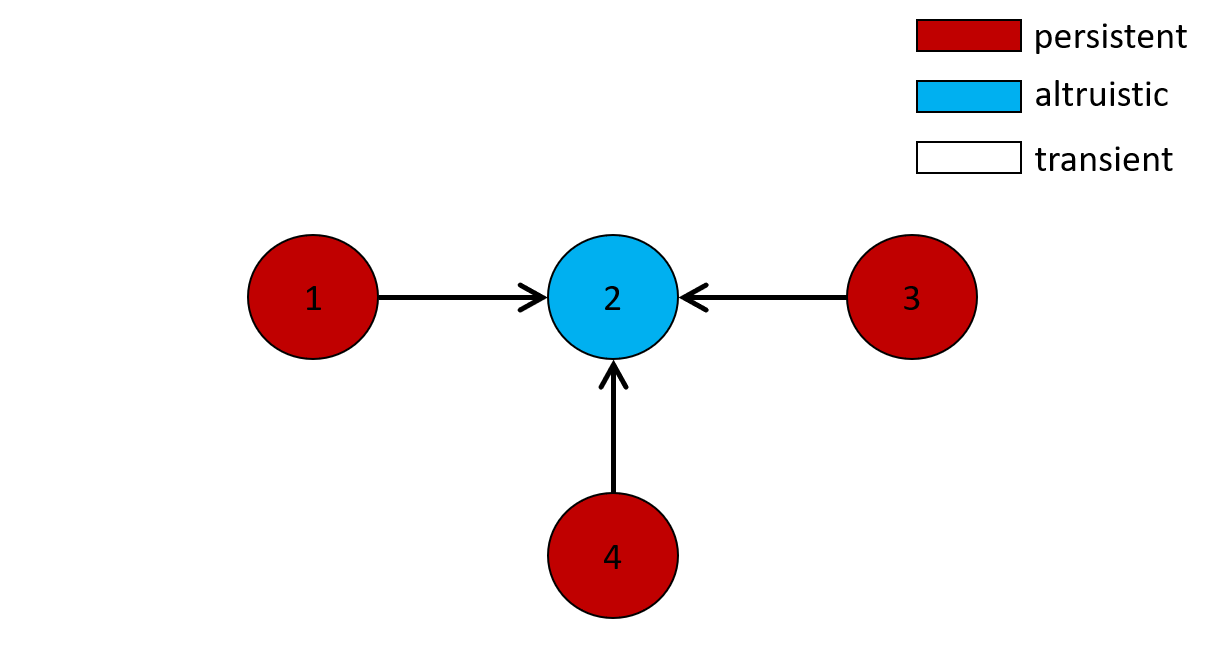}
         \caption{}
         \label{fig:per4_CRN2_fp11}
     \end{subfigure}
        \caption{Stable states where the new viral variant connected to the altruistic variant becomes persistent in a symmetric network}
        \label{fig:per4_CRN2}
\end{figure}

\noindent
When a new viral variant that is cross-immunoreactive with the pre-existing altruistic variant is added to this network, the functions of the viruses could change in two possible ways. The fixed point shown in figure \ref{fig:per4_CRN2_fp2} is described as follows:

\begin{flalign*}
    x_1 &= \frac{b(\beta f_1 + (\alpha - \beta)f_4)}{\beta c p}, &x_2 &= 0, &x_3 &= 0,  &x_4 &= \frac{b f_4(1 - \alpha)}{\beta c p}\\
    r_1 &= \frac{f_1 - f_4}{p}, &r_2 &= \frac{f_4}{\beta p}, &r_3 &= 0, &r_4 &= 0
\end{flalign*}

The stability conditions of this fixed point are $\alpha > \frac{1}{2}, f_1 > f_4, f_4 > f_3,$ and $f_4 > \beta f_2$. Naturally, there exists a symmetric solution, with the variant 1 rather than variant 3 being transient.\\

Another fixed point (Fig. \ref{fig:per4_CRN2_fp11}) is given via the following relations
\begin{flalign*}
    x_1 &= \frac{b(\beta f_1 + (\alpha - \beta)f_4)}{\beta c p}, &x_2 &= 0, &x_3 &= \frac{b(\beta f_3 + (\alpha - \beta)f_4)}{\beta c p},  &x_4 &= \frac{b f_4(1 - 2 \alpha)}{\beta c p}\\
    r_1 &= \frac{f_1 - f_4}{p}, &r_2 &= \frac{f_4}{\beta p}, &r_3 &= \frac{f_3 - f_4}{p}, &r_4 &= 0
\end{flalign*}

Conditions of stability of this fixed point are $\frac{1}{3} < \alpha < \frac{1}{2}, f_1 > f_4, f_3 > f_4,$ and $f_4 > \beta f_2$.\\

In both instances the newly added variant becomes persistent only when it is attached to the altruistic variant. This seems to be a natural result from the perspective of the local immunodeficiency mechanism (see Section \ref{sec:model}). In other aspects, however, the instances describe somewhat different evolutionary phenomena. In the solution depicted on Fig. \ref{fig:per4_CRN2_fp2}, the newly emerged variant substitutes previously persistent variant by virtue of having a higher replication rate, thus providing an example of natural selection action under the local immunodefficiency mechanism. In contrast, for the solution from Fig. \ref{fig:per4_CRN2_fp11} the emerging variant has a lower replication rate than the existing persistent variants. Thus, it neither eliminates these variants nor being eliminated itself, but rather co-exists with them. In this environment, previous persistent variants continue to exist in the same role, although under higher immune pressures and lower population sizes.


Furthermore, the second solution reveals the previously unnoticed phenomenon, whereat a dynamical change in the topology of the cross-immunoreactivity network leads to the emergence of a stable LI in the population, when it previously did not exist. Indeed, the symmetric minimal network on Fig. \ref{fig:symmetricCRN} has the stable LI only under the condition $\alpha > \frac{1}{2}$. The stable state of LI exhibited by the solution on Fig. \ref{fig:per4_CRN2_fp11} exist under the condition $\frac{1}{3} < \alpha < \frac{1}{2}$, which means that with these values of $\alpha$ the initial 3-network did not have a stable LI, but acquired it after addition of a new variant to the network. 

\section{Merging of two cross-immmunoreactivity networks}

In this section we present three cases when merging of two minimal symmetric CR networks lead to the state of stable local immunodeficiency. All other analyzed cases of network merging destroy the stability of this state.

\begin{figure}[ht]
     \centering
     \begin{subfigure}[b]{0.45\textwidth}
         \centering
         \includegraphics[width=\textwidth]{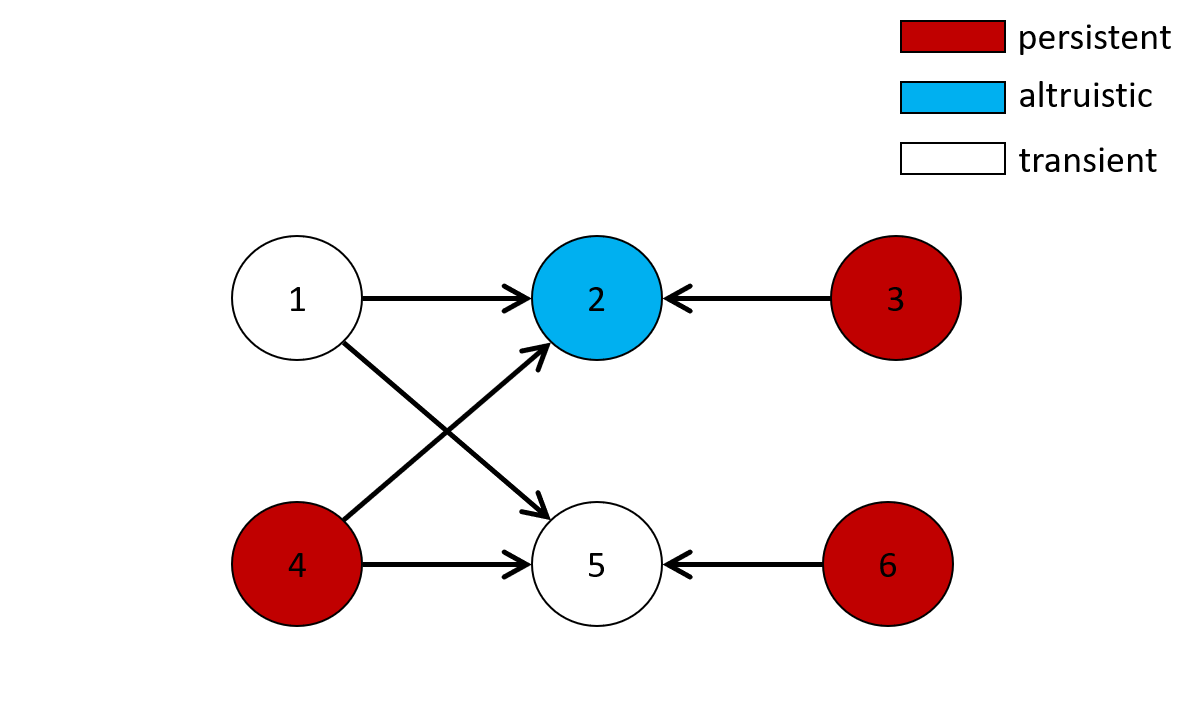}
         \caption{}
         \label{fig:combined_3}
     \end{subfigure}
     \hfill
     \begin{subfigure}[b]{0.45\textwidth}
         \centering
         \includegraphics[width=\textwidth]{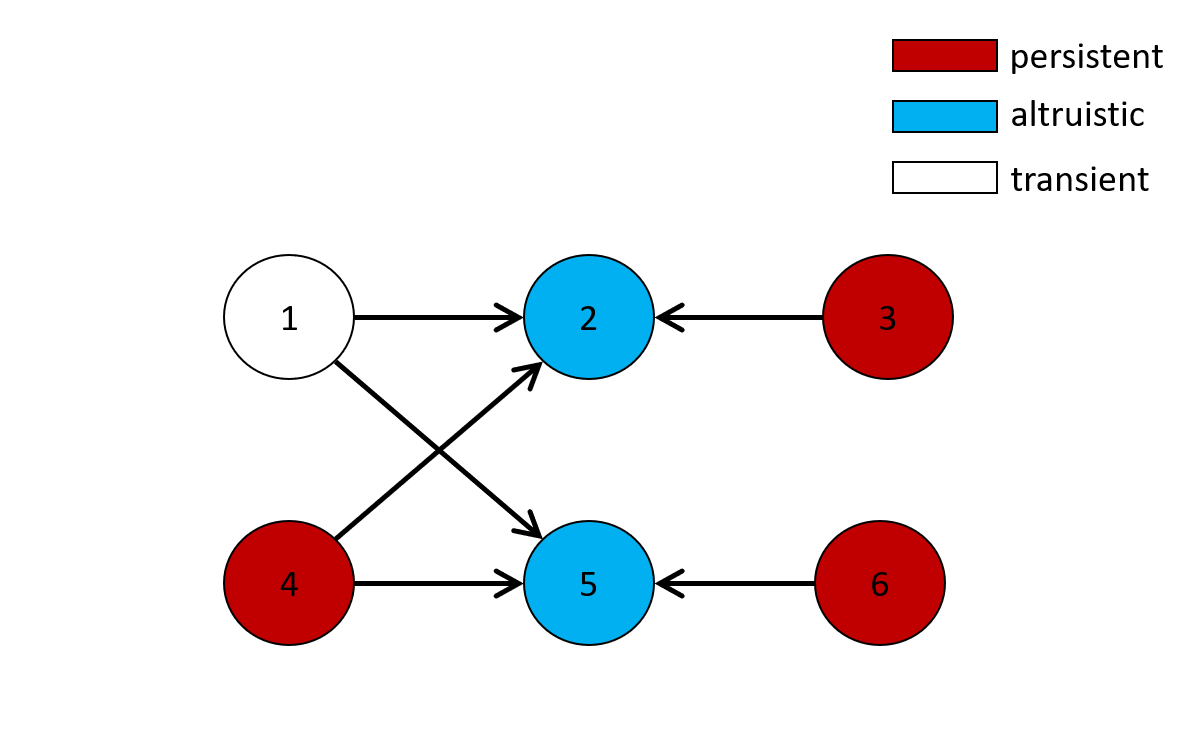}
         \caption{}
         \label{fig:combined_1}
     \end{subfigure}
     \begin{subfigure}[b]{0.45\textwidth}
         \centering
         \includegraphics[width=\textwidth]{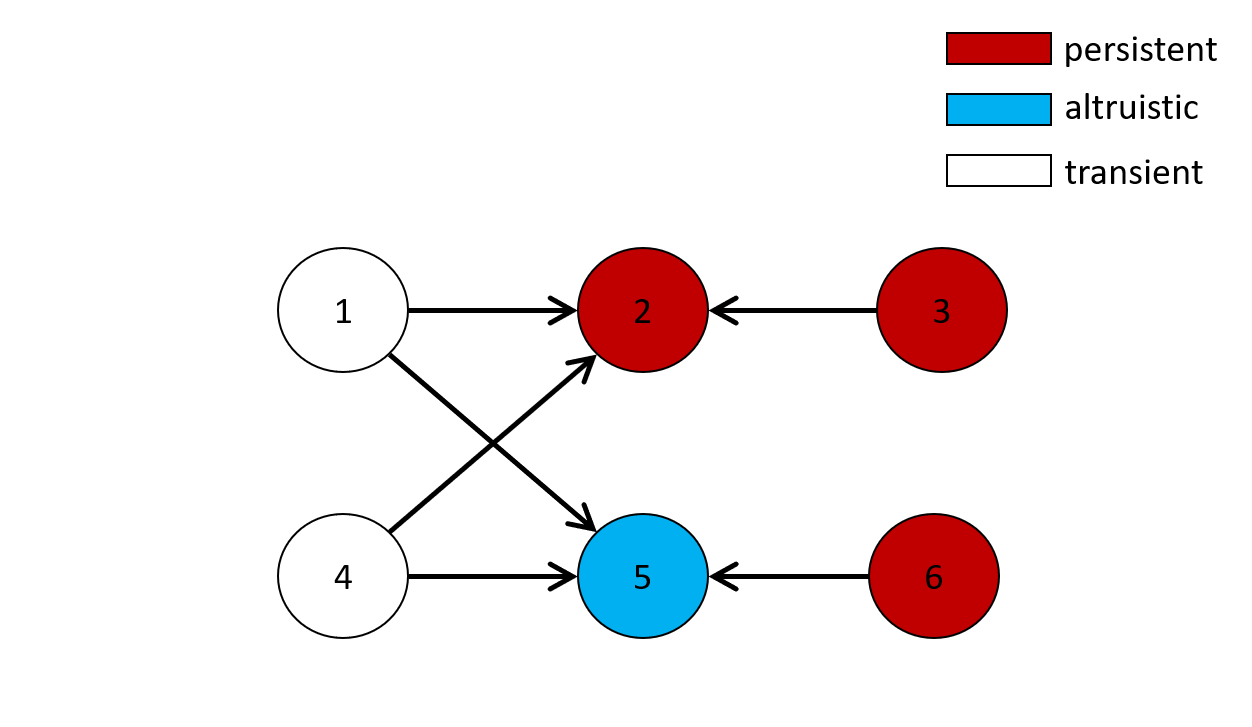}
         \caption{}
         \label{fig:combined_2}
     \end{subfigure}
     \caption{Examples of transformation of function after joining minimal networks}
     \label{fig:combined_networks}
\end{figure}

Fig. \ref{fig:combined_3} depicts a solution for a 6-vertex network obtained by joining two symmetric 3-vertex  networks induced by vertices 1-3 and 4-6, respectively. 
The fixed point corresponding to \ref{fig:combined_3} is given by the following relations
\begin{align*}
    & x_1 = 0, & r_1 &= 0\\
    & x_2 = 0, & r_2 &= \frac{f_4}{\beta p}\\
    & x_3 = \frac{b(\beta f_3 + (\alpha - \beta)f_4)}{\beta c p}, & r_3 &= \frac{f_3 - f_4}{p}\\
    & x_4 = \frac{b f_4(1 - \alpha)}{\beta c p}, & r_4 &= 0\\
    & x_5 = 0, & r_5 &= 0\\
    & x_6 = \frac{b f_6}{c p}, & r_6 &= \frac{f_6}{p}\\
\end{align*}

\noindent
In this solution, previously altruistic variants 5 and previously persistent variant 1 become transient. Once the transient variants are eliminated by the immune system, the cross-immunoreactivity network breaks into two subnetworks, one of which (induced by variants 2,3,4) is isomporphic to a minimal network shown in Fig. \ref{fig:symmetricCRN}. The variant 6 is isolated from the remaining variants, effectively evolves in the absense of cross-immunoreactivity and thus converges to the corresponding stable state.  Another 
possibility leading to the stable state of LI is presented in the Fig. \ref{fig:combined_1}: here a single variant (variant 1) is eliminated, i.e. changes its role from persistent to transient.  The corresponding fixed point of this network is
\begin{align*}
    & x_1 = 0, & r_1 &= 0\\
    & x_2 = 0, & r_2 &= \frac{f_4 - \beta f_5}{\beta p}\\
    & x_3 = \frac{b(\beta f_3 + (\alpha - \beta)(f_4 - \beta f_5)}{\beta c p}, & r_3 &= \frac{f_3 - f_4 + \beta f_5}{p}\\
    & x_4 = \frac{b f_4(1 - \alpha)}{\beta c p}, & r_4 &= 0\\
    & x_5 = 0, & r_5 &= \frac{f_5}{p}\\
    & x_6 = \frac{b ((\alpha - \beta)f_5 + f_6)}{c p}, & r_6 &= \frac{f_6 - \beta f_5}{p}\\
\end{align*}

Finally, the solution from Fig. \ref{fig:combined_2}) describes an outcome, when the elimination 1 and 4 breaks the CR network into two 2-vertex subnetworks reflecting different degrees of local immunodefficiency; both of these states were described in the original publication \cite{skums2015antigenic}. The corresponding fixed point is

\begin{align*}
    & x_1 = 0, & r_1 &= 0\\
    & x_2 = \frac{b f_2 (1 - \alpha)}{c p}, & r_2 &= \frac{f_2}{p}\\
    & x_3 = \frac{b((\alpha - \beta)f_2 + f_3)}{c p}, & r_3 &= \frac{f_3 - \beta f_2}{p}\\
    & x_4 = 0, & r_4 &= 0\\
    & x_5 = 0, & r_5 &= \frac{f_6}{\beta p}\\
    & x_6 = \frac{b f_6}{\beta c p}, & r_6 &= 0\\
\end{align*}

For this solution, the subnetwork induced by vertices 2 and 3 exist in the state, when equilibrium values of $x_3$ and $r_3$ depend not only on $f_3$ but also on $f_2$. It means that the variant $3$ achieves a higher population size under lower immune pressure (in comparison with the system without CR) by exploiting the replicative ability the variant 2. The subnetwork formed by variants 5 and 6 expresses a stronger form of the same phenomenon, where the variant 6 exists without any 5-specific immune pressure (i.e. under the strong state of LI) due to the presence of the 5-specific antibodies, whose high concentration is supported entirely by the variant $6$ (with $r_5$ depending only on $f_6$). The interesting property of the latter subnetwork is that the corresponding subsolution is stable for a positive measure set in the parameter space, when considered within the 6-vertex network; in contrast, it is stable only for $\alpha = 1$, when considered within the 2-vertex network \cite{skums2015antigenic}.

\section{Transformation of functions in evolving networks}

In the previous sections, we analysed the equilibrium solutions describing the asymptotic properties of the system (\ref{eq:mainmodel}). In this section, we discuss the entire dynamics of intra-host viral populations before and after new variants are added to the CRN networks.  Since the model (\ref{eq:mainmodel}) is a highly nonlinear dynamical system, which gives no hope to obtain an analytic solution, the analysis in this section is, by necessity, numerical. In the context of this study, particularly interesting is the speed of transition between different states of the system and change of viral variant roles in the population's intra-host adaptation,  including the elimination of previously persistent variants due to the network expansion.

\begin{figure}[ht]
     \centering
     \begin{subfigure}[b]{0.45\textwidth}
         \centering
         \includegraphics[width=\textwidth]{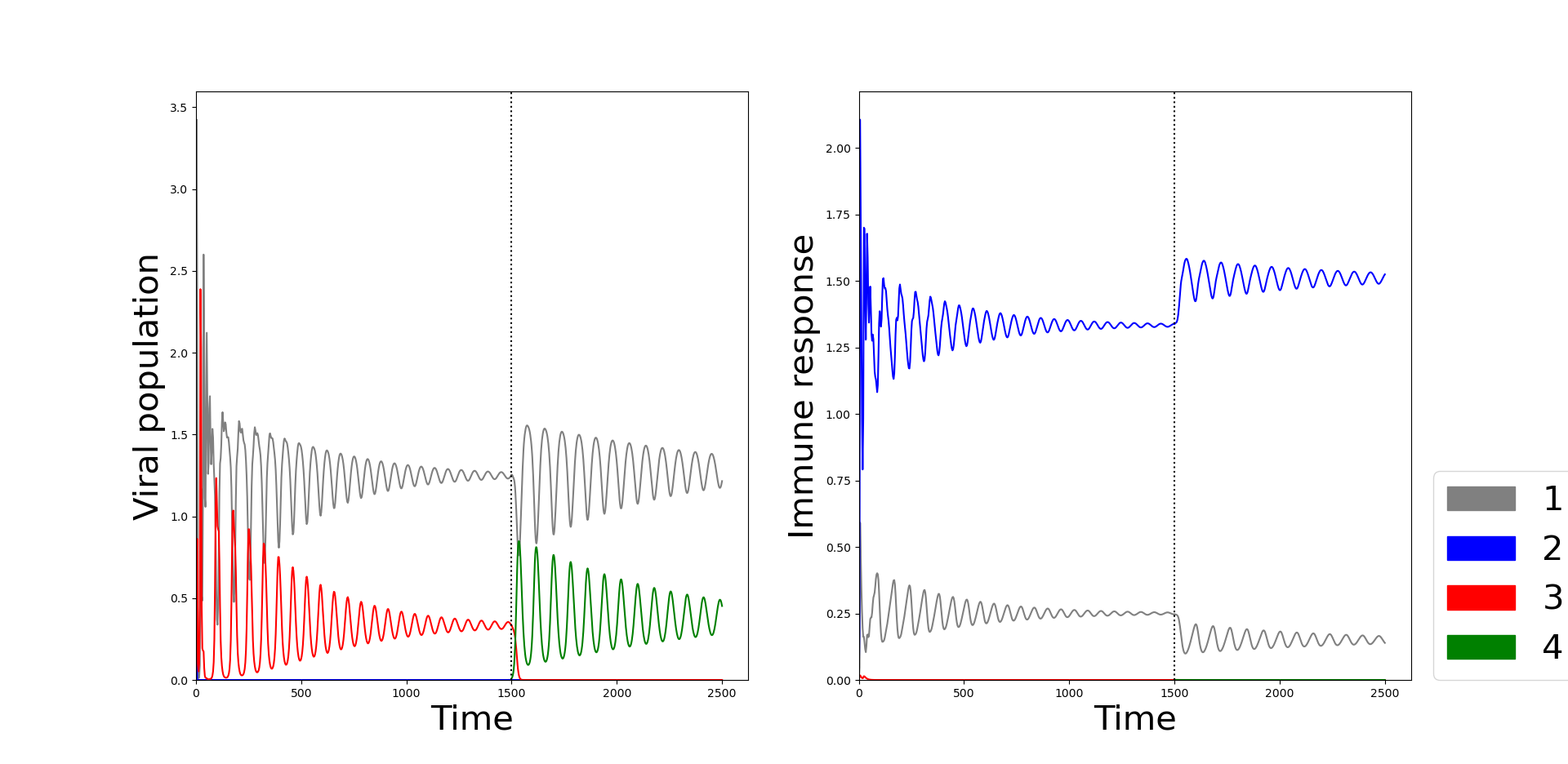}
         \caption{Dynamics of the transformation of function in the network in figure \ref{fig:per4_CRN1_fp2}}
         \label{fig:dynamics_2a}
     \end{subfigure}
     \begin{subfigure}[b]{0.45\textwidth}
         \centering
         \includegraphics[width=\textwidth]{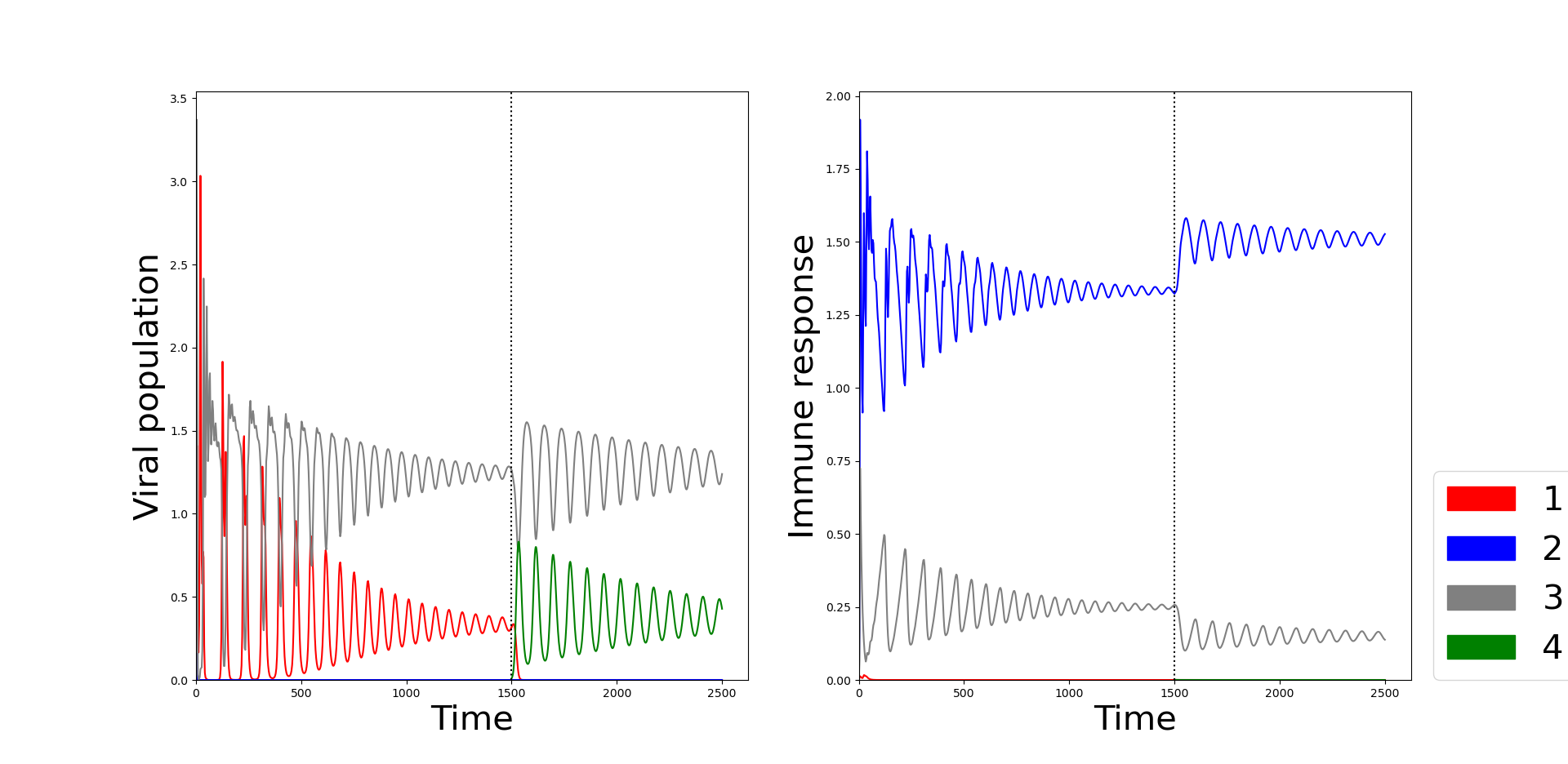}
         \caption{Dynamics of the transformation of function in the network in figure \ref{fig:per4_CRN1_fp3}}
         \label{fig:dynamics_2b}
     \end{subfigure}
     \begin{subfigure}[b]{0.45\textwidth}
         \centering
         \includegraphics[width=\textwidth]{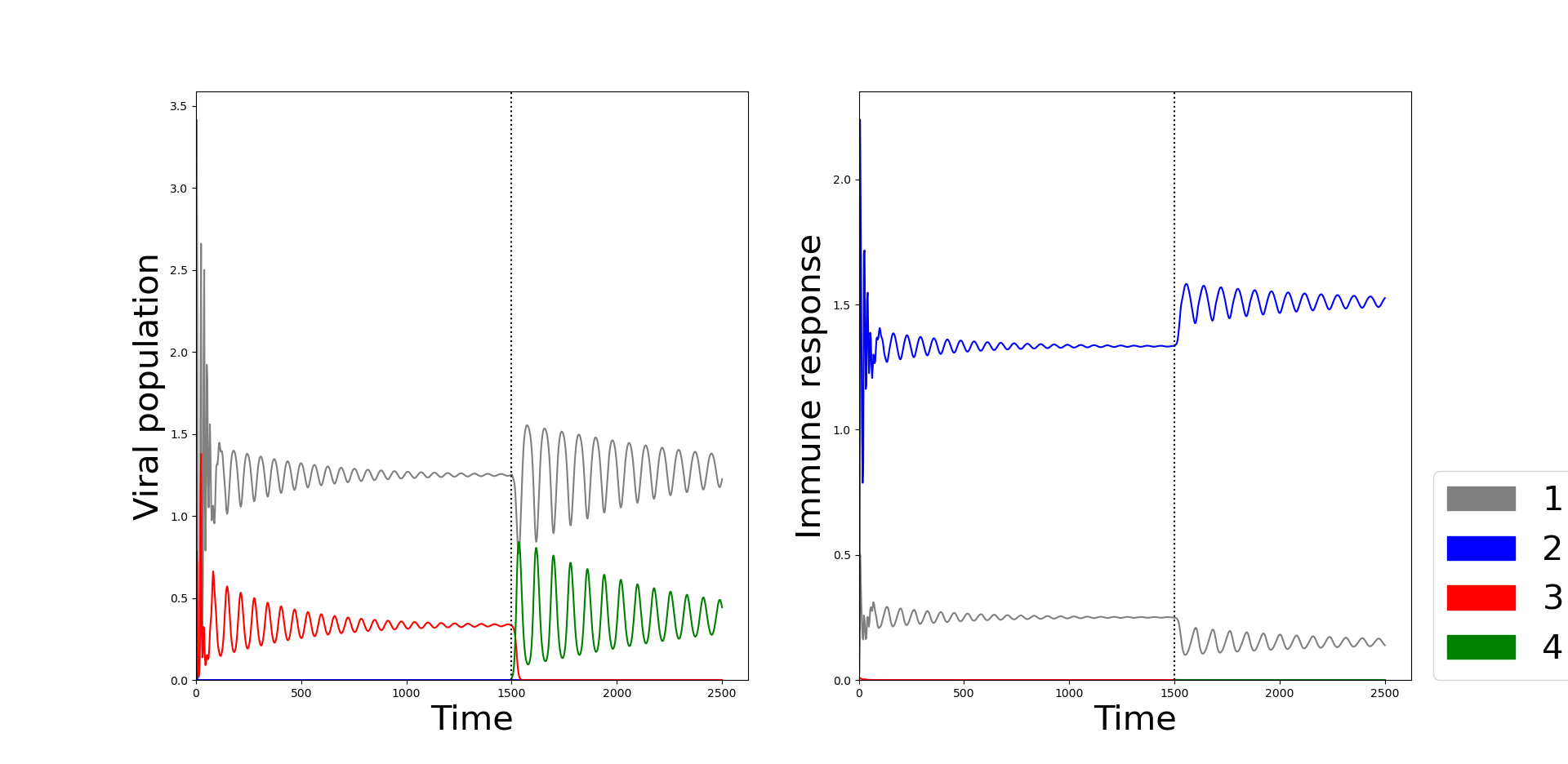}
         \caption{Dynamics of the transformation of function in the network in figure \ref{fig:per4_CRN2_fp2}.}
         \label{fig:dynamics_4a}
     \end{subfigure}
     \begin{subfigure}[b]{0.45\textwidth}
         \centering
         \includegraphics[width=\textwidth]{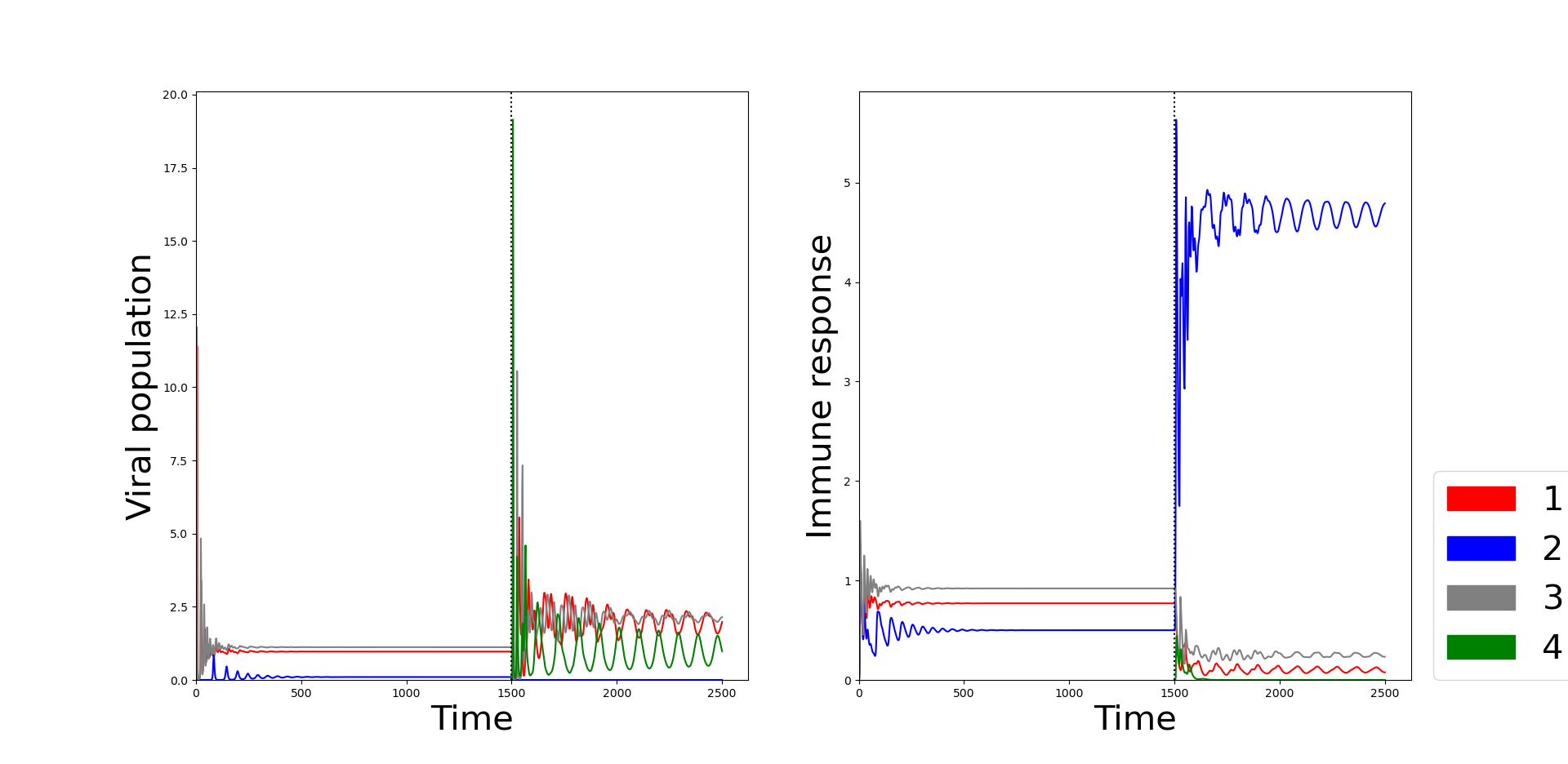}
         \caption{Dynamics of the transformation of function in the network in figure \ref{fig:per4_CRN2_fp11}}
         \label{fig:dynamics_4c}
     \end{subfigure}
        \caption{Dynamics of minimal networks before and after the introduction of a new variant. Left figures depict the dynamics of variant population sizes, and right figures - of the variant-specific immune responses. Dotted lines represent moments of time when the new variants are added to CR networks.}
        \label{fig:dynamics}
\end{figure}


The results are presented on Fig. \ref{fig:dynamics}. Naturally, the dynamics of transformations of the populations from Figures \ref{fig:dynamics_2a} and \ref{fig:dynamics_2b} (Fig. \ref{fig:dynamics_2a}) are qualitatively similar, which is to be expected given the qualitative similarity of their asymptotic solutions (see Subsection \ref{sec:adding1}). In both cases, elimination of previously persistent variants (variants 3 and 1, respectively) happens quite quickly. The same is true for the immune response against the altruistic variant 2, which is boosted by the emergence of a new immunogen 4, thus allowing to sustain the adaptation of two persistent variants (1,4 and 3,4, respectively) under the state of local immunodefficiency.

In contrast, the time evolution of populations shown in figures \ref{fig:dynamics_4a} and \ref{fig:dynamics_4c} essentially differ from each other, with the speed of transition of the latter population being significantly more rapid. As above, this difference can be explained by the properties of the corresponding asymptotic solutions. Indeed, the initial state  in the first network (Fig. \ref{fig:dynamics_4a}) is a stable local immunodeficiency. On the contrary, the initial state in the network in Fig. \ref{fig:dynamics_4c} is unstable local immunodeficiency.  Therefore it is natural that the transition between stable states goes slower. Furthermore, higher concentration of the altruistic variants-specific antibodies achieved for the population  \ref{fig:dynamics_4c} allows to sustain the adaptation of 3 rather than 2 persistent variants.

\section{Conclusions}

In this paper, we study the dynamic and equilibrium properties of a model \cite{skums2015antigenic,bunimovich2019local,bunimovich2020local1,bunimovich2019specialization} describing the behaviour of intra-host viral population that is organized into cross-immunoreactivity (CR) networks and is under pressure by the host's adaptive immune system in the form of variant-specific B-cells. One of the prominent features of this model is the emergence of so-called state of local immunodefficiency, i.e. the equilibrium state where the immune neutralization of certain variants is suppressed due to the interactions between pre-existing antigens and antibodies mediated by the CR network. We concentrate on the transitions between the population states caused by dynamic changes of the CR network topology.  Specifically, we
investigate two events – introduction of a new antigenic variant to the CR network, and the merging of two CR networks in the state of stable immune-adapted equilibrium, which may occur, for example, from a viral transmission between two chronically infected hosts.

It was shown that with the emergence of a new antigenic variant, there can be a rapid rearrangement in the roles played by the variants. A newly emerged variant can become persistent under the following two conditions:  1) the new antigenic variant is cross-immunoreactive with the antibodies specific to the existing altruistic variant and 2) the newly emerged variant has a higher replication rate than a previously persistent variant. This type of rearrangement is expected when the initial system had stable local immunodeficiency before the emergence of a new variant. 

Furthermore, we have shown that the appearance of a novel antigenic variant results in the establishment of a stable local immunodeficiency within a viral population that initially did not exhibit such a condition. This finding diverges from the outcomes of earlier studies, which primarily concentrated on identifying fixed CR networks with a stable local immunodeficiency state. This discovery paves the way for further exploration of CR network dynamics leading to the development of stable LI states.

Similar transitions have been observed for the merging of CR networks. It was shown in several examples how the roles of antigenic variants in the CR networks are rearranged, how certain variants are eliminated, and how the CR networks can break down into subnetworks with different phenotypes.

In addition to the analytical results of equilibrium states, we also analyze numerically the time-evolution of the dynamics of CR networks before and after the emergence of a new viral variant. We find that the transition between two different stable LI states is slower compared to the transition that creates a stable state of LI.

\appendix

\section{Appendix}
Computations corresponding to the minimal networks can be found in \citep{BS1} (branch-cycle network, Fig. \ref{fig:branch_cycle}), and \citep{BS2} (symmetric network, Fig. \ref{fig:symmetricCRN}).

\subsection{Computations for the branch-cycle network with a newly added variant connected to the pre-exsisting altruistic variant}
Fig. \ref{fig:per4_CRN1} depicts two configurations of the branch-cycle network with stable state of local immunodeficiency, where the variant 4 is newly added.\\

The dynamics (\ref{population}) of this population is described by the following equations
\begin{flalign}\label{eqn:per4_CRN1}
\begin{split}
\dot x_1 &= f_1 x_1 - p x_1 (r_1 + \beta r_2),\\
\dot x_2 &= f_2 x2 - p x_2 (r_2 + \beta r_3),\\
\dot x_3 &= f_3 x_3 - p x_3 (r_3 + \beta r_2),\\
\dot x_4 &= f_4 x_4 - p x_4 (r_4 + \beta r_2),\\
\dot r_1 &= c (\frac{x_1 r_1}{r_1 + \alpha r_2}) - b r_1,\\
\dot r_2 &= c (\frac{\alpha x_1 r_2}{r_1 + \alpha r_2} + \frac{x_2 r_2}{r_2 + \alpha r_3} + \frac{\alpha x_3 r_2}{r_3 + \alpha r_2} + \frac{\alpha x_4 r_2}{r_4 + \alpha r_2}) - b r_2,\\
\dot r_3 &= c (\frac{\alpha x_2 r_3}{r_2 + \alpha r_3} + \frac{x_3 r_3}{r_3 + \alpha r_2}) - b r_3,\\
\dot r_4 &= c (\frac{x_4 r_4}{r_4 + \alpha r_2}) - b r_4.
\end{split}
\end{flalign}

The Jacobian of the system of  (\ref{eqn:per4_CRN1}) at the fixed point shown in figure \ref{fig:per4_CRN1_fp2} is:
$$ \scalemath{0.7}{
\left(\begin{array}{cccccccc} 0 & 0 & 0 & 0 & -\frac{b\,\left(\alpha \,f_{4}+\beta \,f_{1}-\beta \,f_{4}\right)}{\beta \,c} & -\frac{b\,\left(\alpha \,f_{4}+\beta \,f_{1}-\beta \,f_{4}\right)}{c} & 0 & 0\\ 0 & f_{2}-\frac{f_{4}}{\beta } & 0 & 0 & 0 & 0 & 0 & 0\\ 0 & 0 & f_{3}-f_{4} & 0 & 0 & 0 & 0 & 0\\ 0 & 0 & 0 & 0 & 0 & \frac{b\,f_{4}\,\left(\alpha -1\right)}{c} & 0 & \frac{b\,f_{4}\,\left(\alpha -1\right)}{\beta \,c}\\ \frac{\beta \,c\,\left(f_{1}-f_{4}\right)}{\alpha \,f_{4}+\beta \,f_{1}-\beta \,f_{4}} & 0 & 0 & 0 & -\frac{b\,\beta \,\left(f_{1}-f_{4}\right)}{\alpha \,f_{4}+\beta \,f_{1}-\beta \,f_{4}} & -\frac{\alpha \,b\,\beta \,\left(f_{1}-f_{4}\right)}{\alpha \,f_{4}+\beta \,f_{1}-\beta \,f_{4}} & 0 & 0\\ \frac{\alpha \,c\,f_{4}}{\alpha \,f_{4}+\beta \,f_{1}-\beta \,f_{4}} & c & c & c & -\frac{\alpha \,b\,f_{4}}{\alpha \,f_{4}+\beta \,f_{1}-\beta \,f_{4}} & -\frac{b\,\left(\alpha \,f_{4}+\beta \,f_{1}-\beta \,f_{4}-\alpha \,\beta \,f_{1}+\alpha \,\beta \,f_{4}\right)}{\alpha \,f_{4}+\beta \,f_{1}-\beta \,f_{4}} & 0 & \frac{b\,\left(\alpha -1\right)}{\alpha }\\ 0 & 0 & 0 & 0 & 0 & 0 & -b & 0\\ 0 & 0 & 0 & 0 & 0 & 0 & 0 & -\frac{b\,\left(2\,\alpha -1\right)}{\alpha } \end{array}\right)
}$$

We will verify stability of this fixed point by analyzing the characteristic polynomial of this Jacobian. Recall at first the conditions of stability of the fixed point, which were mentioned earlier in section 3.1 : $\alpha > \frac{1}{2}, f_1 > f_4, f_4 > f_3,$ and $f_4 > \beta f_2$.\\
Consider now the following positive quantities:
\begin{flalign*}
A &= b \beta (2-\alpha) (f_1 - f_4) + \alpha b f_4 > 0,\\
B &= \frac{b}{\alpha} (2 \alpha - 1) + b + \frac{f_4 - \beta f_2}{\beta} > 0,\\
C &= b \beta (f_1 - f_4) (b (1 - \alpha) + f_1) + \alpha b f_1 f_4 > 0,\\
D &= \frac{b}{\beta} (f_4 - \beta f_2) + \frac{b}{\alpha} (2 \alpha - 1) (b + \frac{f_4 - \beta f_2}{\alpha} > 0,\\
E &= b^2 (1 - \alpha) (f_1 - f_4) (\beta f_1 + \alpha f_4) > 0,\\
F &= b^2 f_4 (1 - \alpha) (f_1 - f_4) (\beta f_1 + (\alpha - \beta) f_4) > 0,\\
G &= \frac{b^2}{\alpha \beta} (2 \alpha - 1) (f_4 - \beta f_2) > 0,\\
H &= \beta f_1 + (\alpha - \beta) f_4 > 0.
\end{flalign*}

The coefficients of the characteristic polynomial of this Jacobian can be expressed in terms of these positive quantities as
\begin{align*}
\begin{split}
\text{coeff}(\lambda ^8) &= 1,\\
\text{coeff}(\lambda ^7) &= \frac{A}{H} + B + (f_4 - f_3),\\
\text{coeff}(\lambda ^6) &= \frac{C}{H} + \frac{A}{H} ((f_4 - f_3) + B) + (f_4 - f_3) B + D,\\
\text{coeff}(\lambda ^5) &= \frac{E}{H} + \frac{C}{H}((f_4 - f_3)+B) + \frac{A}{H}((f_4 - f_3)B + D) + G + (f_4 - f_3)D,\\
\text{coeff}(\lambda ^4) &= \frac{b^2}{\alpha \beta} (2 \alpha - 1)(f_4 - f_3)(f_4 - \beta f_2) + \frac{F}{H} + \frac{E}{H}((f_4 - f_3)+B) \\
&+ \frac{C}{H}((f_4 - f_3)B+D) + \frac{A}{H}(G+(f_4 - f_3)D),\\
\text{coeff}(\lambda ^3) &= \frac{b^2}{\alpha \beta H}(2 \alpha - 1)(f_4 - f_3)(f_4 - \beta f_2)A + \frac{F}{H}(((f_4 - f_3)+B) \\
&+ \frac{E}{H}((f_4 - f_3)B+D) + \frac{C}{H}(G + (f_4 - f_3)D),\\
\text{coeff}(\lambda ^2) &= \frac{b^2}{\alpha \beta H}(2 \alpha - 1)(f_4 - f_3)(f_4 - \beta f_2)C + \frac{F}{H}((f_4 - f_3)B + D) \\
&+ \frac{E}{H}(G + (f_4 - f_3)D),\\
\text{coeff}(\lambda ^1) &= \frac{b^2}{\alpha \beta H}(2 \alpha - 1)(f_4 - f_3)(f_4 - \beta f_2)E + \frac{F}{H}(G + (f_4 - f_3)D),\\
\text{coeff}(\lambda ^0) &= \frac{b^2}{\alpha \beta H}(2 \alpha - 1)(f_4 - f_3)(f_4 - \beta f_2)F.
\end{split}
\end{align*}

Because all the coefficients of the characteristic polynomial are positive, then all eigenvalues of the Jacobian are either real negative numbers or complex numbers with negative real parts. Therefore in this case the state of local immunodeficiency is stable.\\

The Jacobian of the differential equations (\ref{eqn:per4_CRN1}) at the fixed point shown in figure \ref{fig:per4_CRN1_fp3} is:
$$\scalemath{0.7}{
\left(\begin{array}{cccccccc} f_{1}-f_{4} & 0 & 0 & 0 & 0 & 0 & 0 & 0\\ 0 & -\frac{f_{4}-\beta \,f_{2}+\beta ^2\,f_{3}-\beta ^2\,f_{4}}{\beta } & 0 & 0 & 0 & 0 & 0 & 0\\ 0 & 0 & 0 & 0 & 0 & -\frac{b\,\left(\alpha \,f_{4}+\beta \,f_{3}-\beta \,f_{4}\right)}{c} & -\frac{b\,\left(\alpha \,f_{4}+\beta \,f_{3}-\beta \,f_{4}\right)}{\beta \,c} & 0\\ 0 & 0 & 0 & 0 & 0 & \frac{b\,f_{4}\,\left(\alpha -1\right)}{c} & 0 & \frac{b\,f_{4}\,\left(\alpha -1\right)}{\beta \,c}\\ 0 & 0 & 0 & 0 & -b & 0 & 0 & 0\\ c & \frac{c\,f_{4}}{f_{4}+\alpha \,\beta \,f_{3}-\alpha \,\beta \,f_{4}} & \frac{\alpha \,c\,f_{4}}{\alpha \,f_{4}+\beta \,f_{3}-\beta \,f_{4}} & c & 0 & -\frac{b\,\left(\alpha \,f_{4}+\beta \,f_{3}-\beta \,f_{4}-\alpha \,\beta \,f_{3}+\alpha \,\beta \,f_{4}\right)}{\alpha \,f_{4}+\beta \,f_{3}-\beta \,f_{4}} & -\frac{\alpha \,b\,f_{4}}{\alpha \,f_{4}+\beta \,f_{3}-\beta \,f_{4}} & \frac{b\,\left(\alpha -1\right)}{\alpha }\\ 0 & \frac{\alpha \,\beta \,c\,\left(f_{3}-f_{4}\right)}{f_{4}+\alpha \,\beta \,f_{3}-\alpha \,\beta \,f_{4}} & \frac{\beta \,c\,\left(f_{3}-f_{4}\right)}{\alpha \,f_{4}+\beta \,f_{3}-\beta \,f_{4}} & 0 & 0 & -\frac{\alpha \,b\,\beta \,\left(f_{3}-f_{4}\right)}{\alpha \,f_{4}+\beta \,f_{3}-\beta \,f_{4}} & -\frac{b\,\beta \,\left(f_{3}-f_{4}\right)}{\alpha \,f_{4}+\beta \,f_{3}-\beta \,f_{4}} & 0\\ 0 & 0 & 0 & 0 & 0 & 0 & 0 & -\frac{b\,\left(2\,\alpha -1\right)}{\alpha } \end{array}\right)}
$$
Similar to the previous case, it can be shown that this system has the state of stable local immunodeficiency.

\subsection{Computations for the symmetric network with a newly added variant connected to the pre-existing altruistic variant}
Fig. \ref{fig:per4_CRN2} depicts two stable configurations of the branch-cycle network where the variant 4 is newly added.\\

The dynamics (\ref{population}) of this population is described by the following equations
\begin{flalign}\label{eqn:per4_CRN2_1}
\begin{split}
\dot x_1 &= f_1 x_1 - p x_1 (r_1 + \beta r_2),\\
\dot x_2 &= f_2 x2 - p x_2 r_2,\\
\dot x_3 &= f_3 x_3 - p x_3 (r_3 + \beta r_2),\\
\dot x_4 &= f_4 x_4 - p x_4 (r_4 + \beta r_2),\\
\dot r_1 &= c (\frac{x_1 r_1}{r_1 + \alpha r_2}) - b r_1,\\
\dot r_2 &= c (\frac{\alpha x_1 r_2}{r_1 + \alpha r_2} + x_2 + \frac{\alpha x_3 r_2}{r_3 + \alpha r_2} + \frac{\alpha x_4 r_2}{r_4 + \alpha r_2}) - b r_2,\\
\dot r_3 &= c (\frac{x_3 r_3}{r_3 + \alpha r_2}) - b r_3,\\
\dot r_4 &= c (\frac{x_4 r_4}{r_4 + \alpha r_2}) - b r_4.
\end{split}
\end{flalign}

At the fixed point shown in the figure \ref{fig:per4_CRN2_fp2}, the Jacobian of the system  (\ref{eqn:per4_CRN2_1}) equals
$$\scalemath{0.7}{
\left(\begin{array}{cccccccc} 0 & 0 & 0 & 0 & -\frac{b\,\left(\alpha \,f_{4}+\beta \,f_{1}-\beta \,f_{4}\right)}{\beta \,c} & -\frac{b\,\left(\alpha \,f_{4}+\beta \,f_{1}-\beta \,f_{4}\right)}{c} & 0 & 0\\ 0 & f_{2}-\frac{f_{4}}{\beta } & 0 & 0 & 0 & 0 & 0 & 0\\ 0 & 0 & f_{3}-f_{4} & 0 & 0 & 0 & 0 & 0\\ 0 & 0 & 0 & 0 & 0 & \frac{b\,f_{4}\,\left(\alpha -1\right)}{c} & 0 & \frac{b\,f_{4}\,\left(\alpha -1\right)}{\beta \,c}\\ \frac{\beta \,c\,\left(f_{1}-f_{4}\right)}{\alpha \,f_{4}+\beta \,f_{1}-\beta \,f_{4}} & 0 & 0 & 0 & -\frac{b\,\beta \,\left(f_{1}-f_{4}\right)}{\alpha \,f_{4}+\beta \,f_{1}-\beta \,f_{4}} & -\frac{\alpha \,b\,\beta \,\left(f_{1}-f_{4}\right)}{\alpha \,f_{4}+\beta \,f_{1}-\beta \,f_{4}} & 0 & 0\\ \frac{\alpha \,c\,f_{4}}{\alpha \,f_{4}+\beta \,f_{1}-\beta \,f_{4}} & c & c & c & -\frac{\alpha \,b\,f_{4}}{\alpha \,f_{4}+\beta \,f_{1}-\beta \,f_{4}} & -\frac{b\,\left(\alpha \,f_{4}+\beta \,f_{1}-\beta \,f_{4}-\alpha \,\beta \,f_{1}+\alpha \,\beta \,f_{4}\right)}{\alpha \,f_{4}+\beta \,f_{1}-\beta \,f_{4}} & 0 & \frac{b\,\left(\alpha -1\right)}{\alpha }\\ 0 & 0 & 0 & 0 & 0 & 0 & -b & 0\\ 0 & 0 & 0 & 0 & 0 & 0 & 0 & -\frac{b\,\left(2\,\alpha -1\right)}{\alpha } \end{array}\right)
}$$

Again, consider characteristic polynomial of this Jacobian. The conditions of stability of the fixed point are (see the section 3.1) $\alpha > \frac{1}{2}, f_1 > f_4, f_4 > f_3,$ and $f_4 > \beta f_2$.\\
The following quantities are all positive:
\begin{flalign*}
A &= b \beta (2-\alpha) (f_1 - f_4) + \alpha b f_4 > 0,\\
B &= b \beta (f_1 - f_4) (b (1 - \alpha) + f_1) + \alpha b f_1 f_4 > 0,\\
C &= b^2 (1 - \alpha) (f_1 - f_4) (\beta f_1 + \alpha f_4) > 0,\\
D &= b^2 f_4 (1 - \alpha) (f_1 - f_4) (\beta f_1 + (\alpha - \beta) f_4) > 0,\\
E &= \frac{(f_4 - f_3) + b(2\alpha-1)/\alpha + b + (f_4 - \beta f_2)/\beta}{\beta f_1 + (\alpha - \beta) f_4}  > 0,\\
F &= \frac{(f_4 - f_3)(\frac{\beta}{\alpha}(2\alpha - 1) + b + \frac{f_4-\beta f_2}{\beta})+\frac{b}{\beta}(f_4 - \beta f_2) + b(2\alpha - 1)(\frac{b}{\alpha} + \frac{f_4 - \beta f_2}{\alpha \beta})}{\beta f_1 + (\alpha - \beta) f_4} > 0,\\
G &= \frac{\frac{b^2}{\alpha\beta}(2\alpha-1)(f_4-\beta f_2) + (f_4-f_3)(\frac{b}{\beta}(f_4 - \beta f_2) + b(a\alpha-1)(\frac{b}{\alpha}+\frac{f_4 - \beta f_2}{\alpha \beta}))}{\beta f_1 + (\alpha - \beta) f_4} > 0,\\
H &= \beta f_1 + (\alpha - \beta) f_4 > 0.
\end{flalign*}
The coefficients of the characteristic polynomial of this Jacobian can be expressed in terms of these positive quantities as
\begin{align*}
\begin{split}
\text{coeff}(\lambda ^8) &= 1,\\
\text{coeff}(\lambda ^7) &= \frac{A}{H} + EH,\\
\text{coeff}(\lambda ^6) &= \frac{B}{H} + AE + FH,\\
\text{coeff}(\lambda ^5) &= \frac{C}{H} + BE + AF + GH,\\
\text{coeff}(\lambda ^4) &= \frac{b^2}{\alpha \beta} (2 \alpha - 1)(f_4 - f_3)(f_4 - \beta f_2) + \frac{D}{H} + CE + BF + AG,\\
\text{coeff}(\lambda ^3) &= \frac{b^2}{\alpha \beta H}(2 \alpha - 1)(f_4 - f_3)(f_4 - \beta f_2)A + DE + CF + BG,\\
\text{coeff}(\lambda ^2) &= \frac{b^2}{\alpha \beta H}(2 \alpha - 1)(f_4 - f_3)(f_4 - \beta f_2)A + DE + CF + BG,\\
\text{coeff}(\lambda ^1) &= \frac{b^2}{\alpha \beta H}(2 \alpha - 1)(f_4 - f_3)(f_4 - \beta f_2)C + DG,\\
\text{coeff}(\lambda ^0) &= \frac{b^4 f_4}{\alpha \beta}(1-\alpha)(2 \alpha - 1)(f_1 - f_4)(f_4 - f_3)(f_4 - \beta f_2).
\end{split}
\end{align*}
Because the coefficients of the characteristic polynomial are positive, then all eigenvalues of the Jacobian are either real negative or complex with negative real parts. Therefore in this  viral network the state of local immunodeficiency is stable.\\

At the fixed point of the network in the figure \ref{fig:per4_CRN2_fp11}, the Jacobian of the system of equations (\ref{eqn:per4_CRN2_1}) is $J = (X Y Z)$ where
$$
X = 
\left(\begin{array}{ccccc} 0 & 0 & 0 & 0 & -\frac{b\,\left(\alpha \,f_{4}+\beta \,f_{1}-\beta \,f_{4}\right)}{\beta \,c}\\ 0 & f_{2}-\frac{f_{4}}{\beta } & 0 & 0 & 0\\ 0 & 0 & 0 & 0 & 0\\ 0 & 0 & 0 & 0 & 0\\ \frac{\beta \,c\,\left(f_{1}-f_{4}\right)}{\alpha \,f_{4}+\beta \,f_{1}-\beta \,f_{4}} & 0 & 0 & 0 & -\frac{b\,\beta \,\left(f_{1}-f_{4}\right)}{\alpha \,f_{4}+\beta \,f_{1}-\beta \,f_{4}}\\ \frac{\alpha \,c\,f_{4}}{\alpha \,f_{4}+\beta \,f_{1}-\beta \,f_{4}} & c & \frac{\alpha \,c\,f_{4}}{\alpha \,f_{4}+\beta \,f_{3}-\beta \,f_{4}} & c & -\frac{\alpha \,b\,f_{4}}{\alpha \,f_{4}+\beta \,f_{1}-\beta \,f_{4}}\\ 0 & 0 & \frac{\beta \,c\,\left(f_{3}-f_{4}\right)}{\alpha \,f_{4}+\beta \,f_{3}-\beta \,f_{4}} & 0 & 0\\ 0 & 0 & 0 & 0 & 0 \end{array}\right),
$$
\begin{equation*}
  Y = \scalemath{0.7}{
\begin{pmatrix} 
-\frac{b\,\left(\alpha \,f_{4}+\beta \,f_{1}-\beta \,f_{4}\right)}{c}\\
0\\
-\frac{b\,\left(\alpha \,f_{4}+\beta \,f_{3}-\beta \,f_{4}\right)}{c}\\ \frac{b\,f_{4}\,\left(2\,\alpha -1\right)}{c}\\
-\frac{\alpha \,b\,\beta \,\left(f_{1}-f_{4}\right)}{\alpha \,f_{4}+\beta \,f_{1}-\beta \,f_{4}}\\
\begin{aligned}
\Bigg(
\frac{1}{(\beta f_1 + (\alpha - \beta)f_4)\,(\beta f_3 + (\alpha - \beta)f_4)}
(b\,(\alpha ^2\,{f_{4}}^2+\beta ^2\,{f_{4}}^2-2\,\alpha \,\beta ^2\,{f_{4}}^2+2\,\alpha ^2\,\beta \,{f_{4}}^2 & \\[-1.2ex] 
-2\,\alpha \,\beta \,{f_{4}}^2 +\beta ^2\,f_{1}\,f_{3}-\beta ^2\,f_{1}\,f_{4}-\beta ^2\,f_{3}\,f_{4}-2\,\alpha \,\beta ^2\,f_{1}\,f_{3}+ & \\ 
2\,\alpha \,\beta ^2\,f_{1}\,f_{4}- \alpha ^2\,\beta \,f_{1}\,f_{4} +2\,\alpha \,\beta ^2\,f_{3}\,f_{4}-\alpha ^2\,\beta \,f_{3}\,f_{4}+\alpha \,\beta \,f_{1}\,f_{4}+\alpha \,\beta \,f_{3}\,f_{4})) 
\Bigg)\\
\end{aligned}\\
 -\frac{\alpha \,b\,\beta \,\left(f_{3}-f_{4}\right)}{\alpha \,f_{4}+\beta \,f_{3}-\beta \,f_{4}}\\
0
\end{pmatrix}  }
\end{equation*}
and
$$
Z = 
\left(\begin{array}{cc} 0 & 0\\ 0 & 0\\ -\frac{b\,\left(\alpha \,f_{4}+\beta \,f_{3}-\beta \,f_{4}\right)}{\beta \,c} & 0\\ 0 & \frac{b\,f_{4}\,\left(2\,\alpha -1\right)}{\beta \,c}\\ 0 & 0\\ -\frac{\alpha \,b\,f_{4}}{\alpha \,f_{4}+\beta \,f_{3}-\beta \,f_{4}} & \frac{b\,\left(2\,\alpha -1\right)}{\alpha }\\ -\frac{b\,\beta \,\left(f_{3}-f_{4}\right)}{\alpha \,f_{4}+\beta \,f_{3}-\beta \,f_{4}} & 0\\ 0 & -\frac{b\,\left(3\,\alpha -1\right)}{\alpha } \end{array}\right).
$$

As mentioned above, this is a rather unique fixed point. It takes also lengthier computations to prove its stability. Recall the conditions for stability of this system, i.e. $\frac{1}{3} < \alpha < \frac{1}{2}, f_1 > f_4, f_3 > f_4,$ and $f_4 > \beta f_2$.\\
Now consider the positive quantities:
\begin{flalign*}
\begin{split}
    A &= b\beta (3\alpha-1) + \alpha(f_1 - \beta f_2) > 0,\\
    B &= \alpha^2 f_1^2 + (f_3-f_1)(\alpha\beta f_1(2-\alpha)+\beta^2(f_4-f_1)+2\beta^2(f_4-f_1)(1-\alpha)) \\
    &+\alpha\beta f_1(2-\alpha)(f_4-f_1) > 0,\\
    C &= (f_3-f_1)(\alpha \beta b f_1 (1-\alpha)+f_1^2(\alpha^2+\beta^2)+\beta f_1 f_3 (\alpha-\beta) + \beta^2 f_4(f_3 - f_1)\\
    &+b \beta^2 (3-4\alpha)(f_4-f_1)) + 
    (f_4-f_1)(\alpha \beta b f_1 (1-\alpha) + \beta(2\alpha f_1 + \beta f_4)(f_3-f_1)\\
    &+2\alpha\beta f_1 f_3) + \alpha^2 f_1^2 f_4 > 0,\\
    D &= (1-\alpha)(f_1(\alpha-\beta)((f_3-f_1)(f_1(\alpha-\beta)+\beta f_3 + 2\beta f_4) + (f_4 - f_1)(f_1 (\alpha-\beta)\\
    &+\beta f_4 + 2\beta f_3))+ \beta^2 (f_3 + f_4)(f_3(f_4-f_1)+f_4(f_3-f_1)) > 0,\\
    E &= (1-2\alpha)((f_1-f_3)(b \beta f_1 (\alpha-\beta)(\alpha f_1 -2f_4) + b\beta^2 (f_4(f_3-f_4) + f_1 f_3) \\
    &- f_1^4(\alpha-\beta)^2)) + (\alpha-\beta)f_1(\alpha\beta f_1 f_3(f_1-f_4) - \beta f_1^2(1-2\alpha)(f_3+f_4)  \\
    &+ \alpha\beta f_1 f_4(f_1-f_3)- \alpha^2 f_1 f_3 - \alpha \beta f_1 f_3^2 - \alpha^2 f_1^2 f_4 + \alpha f_3 f_4 (f_1 + f_3) - \alpha\beta f_1 f_4^2 \\
    & + \beta f_3 f_4^2) + f_3 f_4 \beta (\alpha\beta f_1((f_1-f_3)+(f_1-f_4))+\beta f_3 f_4 - \alpha f_1^2  > 0,\\
    F &= ((\alpha - \beta)f_1 + \beta f_3)((\alpha - \beta)f_1 + \beta f_4) > 0.
\end{split}
\end{flalign*}

Note that some of the quantities in these calculations were found to be positive through direct numerical computations. The coefficients of the characteristic polynomial of this Jacobian can be expressed in terms of these positive quantities as
\begin{align*}
\begin{split}
\text{coeff}(\lambda ^8) &= \frac{f_1^2(\alpha-\beta)^2 + \beta (\alpha-\beta)(f_1 f_3 + f_1 f_4) + \beta^2 f_3 f_4}{((\alpha-\beta)f_1 + \beta f_3)((\alpha - \beta)f_1 + \beta f_4)},\\
\text{coeff}(\lambda ^7) &= \frac{A}{\alpha\beta} + \frac{bB}{F},\\
\text{coeff}(\lambda ^6) &= \frac{(3\alpha-1)b(f_1-\beta f_2)}{\alpha\beta} + \frac{bAB}{\alpha\beta F} + \frac{bC}{F},\\
\text{coeff}(\lambda ^5) &= \frac{(3\alpha-1)b^2(f_1-\beta f_2)B}{\alpha\beta} + \frac{bAC}{\alpha\beta F} + \frac{bD}{F},\\
\text{coeff}(\lambda ^4) &= \frac{(3\alpha-1)b^2(f_1-\beta f_2)C}{\alpha\beta F} + \frac{b^2AD}{\alpha \beta F}  + \frac{b^2 E}{F},\\
\text{coeff}(\lambda ^3) &= \frac{(1-2\alpha)b^3(f_3-f_1)(f_4-f_1)((\alpha^2 - \beta^2)f_1^2 + \alpha\beta f_1f_3 + \alpha\beta f_1 f_4 + \beta^2f_3 f_4)}{F} \\
&+ \frac{(3\alpha-1)b^3(f_1-\beta f_2)D}{\alpha\beta F} + \frac{b^2AE}{\alpha\beta F},\\
\text{coeff}(\lambda ^2) &= (1-2\alpha)b^3 f_1 (f_3-f_1)(f_4-f_1) \\
&+ \frac{(1-2\alpha)b^3 A (f_3-f_1)(f_4-f_1)((\alpha^2 - \beta^2)f_1^2 + \alpha\beta f_1(f_3 + f_4) + \beta^2 f_3 f_4)}{\alpha\beta F} \\
&+ \frac{(3\alpha-1)b^3(f_1-\beta f_2)E}{\alpha\beta F},\\
\text{coeff}(\lambda ^1) &= \frac{1}{\alpha\beta F}( (1-2\alpha)b^3(f_3-f_1)(f_4-f_1)((f_1-\beta f_2)(\alpha^2bf_1^2 + \alpha\beta b f_1f_4 \\
&+ \alpha \beta b f_1 f_3 + b\beta^2 f_3f_4) + (f_3-f_1)b\beta^3f_1(f_4-f_1) + b\beta f_1^2(\alpha^2 f_1 + \beta^2f_2) \\
&+ \alpha\beta^2 bf_1^2 (f_3+f_4)) + (1-\alpha)b\beta^2f_1^3 + (\alpha-\beta)((\alpha-\beta)\alpha f_1^3(f_1-\beta f_2) \\
&+ (f_1 - \beta f_2)\alpha\beta f_1^2(f_3+f_4)) - 6\alpha^2b\beta^2f_1^3),\\
\text{coeff}(\lambda ^0) &= \frac{b^4 f_1(1-2\alpha)(3\alpha-1)(f_1-\beta f_2)(f_1-f_3)(f_1-f_4)}{\alpha\beta}.
\end{split}
\end{align*}

All the coefficients of the characteristic polynomial are positive. Hence all eigenvalues of the Jacobian are either real negative or complex with negative real parts. Therefore in this model the state of local immunodeficiency is stable.\\

\subsection{Computations for minimal networks with a newly added viral variant connected to pre-existing persistent variants} \label{appendix3}
In section 3.1, we noted that the newly emerging variant has to elicit immune response against the pre-existing altruistic variant to maintain stable LI. Here we show the results of computations for cases when the newly emerged variant is connected to either of the two other persistent variants.

When the newly emerged variant is connected to the variant 1, the dynamics (\ref{population}) of this CRN is described by the following equations
\begin{flalign}\label{eqn:per4_CRN2_2}
\begin{split}
\dot x_1 &= f_1 x_1 - p x_1 (r_1 + \beta r_2),\\
\dot x_2 &= f_2 x2 - p x_2 (r_2 + \beta r_3),\\
\dot x_3 &= f_3 x_3 - p x_3 (r_3 + \beta r_2),\\
\dot x_4 &= f_4 x_4 - p x_4 (r_4 + \beta r_1),\\
\dot r_1 &= c (\frac{x_1 r_1}{r_1 + \alpha r_2} + \frac{\alpha x_4 r_1}{\alpha r_1 + r_4}) - b r_1,\\
\dot r_2 &= c (\frac{\alpha x_1 r_2}{r_1 + \alpha r_2} + \frac{x_2 r_2}{r_2 + \alpha r_3} + \frac{\alpha x_3 r_2}{r_3 + \alpha r_2} - b r_2,\\
\dot r_3 &= c (\frac{\alpha x_2 r_3}{r_2 + \alpha r_3} + \frac{x_3 r_3}{r_3 + \alpha r_2}) - b r_3,\\
\dot r_4 &= c (\frac{x_4 r_4}{r_4 + \alpha r_1}) - b r_4.
\end{split}
\end{flalign}

\begin{figure}[ht]
    \centering
    \includegraphics[scale=0.4]{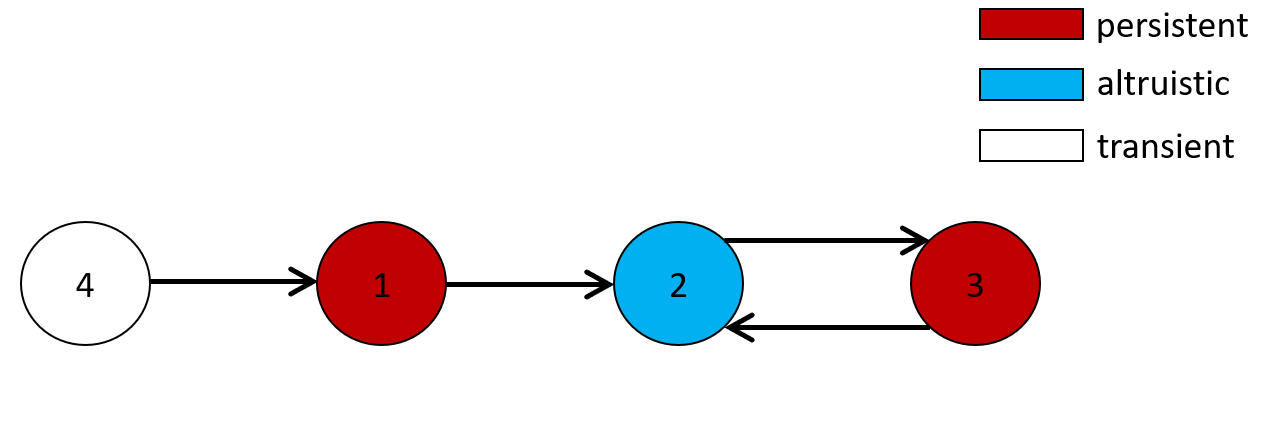}
    \caption{A fixed point when the newly emerging variant is connected to the variant 1 and becomes transient}
    \label{fig:branch_cycle_4to1_inactive}
\end{figure}

In the fixed point shown in figure \ref{fig:branch_cycle_4to1_inactive}, the newly emergent variant becomes transient, and the resulting network is functionally the same as the branch-cycle network (Fig.\ref{fig:branch_cycle}). There is no transformation of functions in this case.

\begin{figure}[ht]
    \centering
    \includegraphics[scale=0.4]{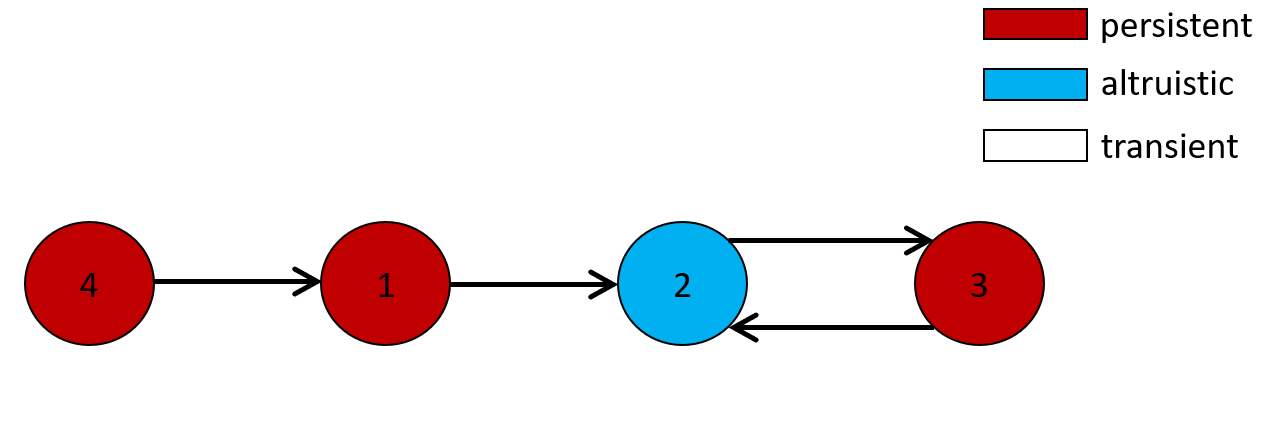}
    \caption{A fixed point when the newly emerging variant is connected to the variant 1 and becomes persistent}
    \label{fig:branch_cycle_4to1_persistent}
\end{figure}
Figure \ref{fig:branch_cycle_4to1_persistent} shows the fixed point
\begin{flalign*}
    x_1 &= \frac{b f_1 (1 - \alpha)}{\beta c p}, &x_2 &= 0, &x_3 &= \frac{b(\beta f_3 + (\alpha - \beta)f_1)}{\beta c p},  &x_4 &= \frac{b f_4}{c p}\\
    r_1 &= 0, &r_2 &= \frac{f_1}{\beta p}, &r_3 &= \frac{f_3 - f_1}{p}, &r_4 &= \frac{f_4}{p}
\end{flalign*}
The Jacobian at this fixed point is 
$$\scalemath{0.7}{
\left(\begin{array}{cccccccc} 
0 & 0 & 0 & 0 & \frac{b f_1 (\alpha -1)}{\beta c} & \frac{b f_1(\alpha -1)}{c} & 0 & 0 \\
0 & -\frac{1}{\beta}(f_1 - \beta f_2 - \beta^2 f_1 + \beta^2 f_3) & 0 & 0 & 0 & 0 & 0 & 0\\ 
0 & 0 & 0 & 0 & 0 & -\frac{b}{c}(\alpha f_1 - \beta f_1 + \beta f_3) & -\frac{b((\alpha - \beta)f_1 + \beta f_3)}{\beta c} & 0\\ 
0 & 0 & 0 & 0 & -\frac{b \beta f_4}{c} & 0 & 0 & -\frac{b f_4}{c}\\ 
0 & 0 & 0 & 0 & \frac{b}{\alpha} (1-\alpha)^2 & 0 & 0 & 0\\ 
c & \frac{c f_1}{f_1 - \alpha \beta f_1 + \alpha \beta f_3} & \frac{\alpha c f_1}{(\alpha - \beta)f_1 + \beta f_3} & 0 & b(1 - \frac{1}{\alpha}) & \frac{-b ((\alpha - \beta)f_1 + \beta f_3 + \alpha \beta (f_1 - f_3))}{(\alpha - \beta)f_1 + \beta f_3} & -\frac{\alpha b f_1}{(\alpha-\beta)f_1 + \beta f_3} & 0\\
0 & \frac{\alpha \beta c(f_3-f_1)}{f_1 - \alpha \beta f_1 + \alpha \beta f_3} & \frac{\beta c(f_3-f_1)}{(\alpha-\beta)f_1 + \beta f_3} & 0 & 0 & \frac{\alpha \beta b(f_1-f_3)}{(\alpha-\beta)f_1 + \beta f_3} & \frac{b\beta(f_1-f_3)}{(\alpha-\beta)f_1 + \beta f_3} & 0\\ 
0 & 0 & 0 & c & -\alpha b & 0 & 0 & -b \end{array}\right)
}$$
This fixed point can exist under the conditions $\alpha < 1, f_1 < \frac{\beta f_3}{\beta - \alpha}$, and $f_3>f_1$.\\
Analyzing the eigenvalues of the system at this point, we get $\lambda_3 = b(\alpha + \frac{1}{\alpha} - 2) > 0$. As at least one of the eigenvalues are positive, the fixed point is unstable. Other fixed points where virus 4 becomes persistent can be similarly proven to have unstable LI.

\begin{figure}[ht]
    \centering
    \includegraphics[scale=0.4]{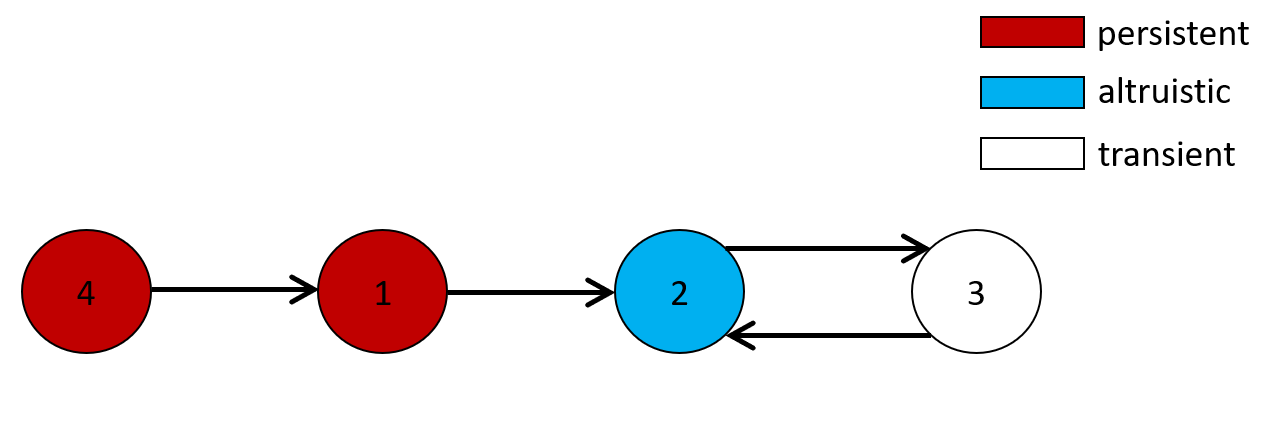}
    \caption{A fixed point when the newly emerging variant is connected to the variant 1 and becomes persistent, making the variant 3 transient}
    \label{fig:branch_cycle_4to1_3inactive}
\end{figure}
The fixed points resulting in the network in figure \ref{fig:branch_cycle_4to1_3inactive} are
\begin{flalign*}
    x_1 &= \frac{b f_1}{\beta c p}, &x_2 &= 0, &x_3 &= 0,  &x_4 &= \frac{b f_4}{c p}\\
    r_1 &= 0, &r_2 &= \frac{f_1}{\beta p}, &r_3 &= 0, &r_4 &= \frac{f_4}{p}
\end{flalign*}
The Jacobian at this fixed point is
$$\scalemath{0.7}{
\left(\begin{array}{cccccccc} 
0 & 0 & 0 & 0 & -\frac{b f_1}{\beta c} & -\frac{b f_1}{c} & 0 & 0 \\
0 & f_2 - \frac{f_1}{\beta} & 0 & 0 & 0 & 0 & 0 & 0\\ 
0 & 0 & f_3 - f_1 & 0 & 0 & 0 & 0 & 0\\ 
0 & 0 & 0 & 0 & -\frac{b \beta f_4}{c} & 0 & 0 & -\frac{b f_4}{c}\\ 
0 & 0 & 0 & 0 & b(\alpha + \frac{1}{\alpha} - 1) & 0 & 0 & 0\\ 
c & c & c & 0 & -\frac{b}{\alpha} & -b & 0 & 0\\
0 & 0 & 0 & 0 & 0 & 0 & -b & 0\\ 
0 & 0 & 0 & c & -\alpha b & 0 & 0 & -b \end{array}\right)
}$$
Analyzing the eigenvalues of the system at this point, we get $\lambda_5 = b(\alpha + \frac{1}{\alpha} - 1) > 0$. As at least one of the eigenvalues are positive, the fixed point is unstable. Other fixed points where virus 4 becomes persistent can be similarly proven to have unstable LI.

Similarly, all networks where virus 4 is connected to virus 1 in the branch-cycle network can be shown to have either no LI or no stable LI.\\

Now we look at cases where the newly emerged variant is connected to the variant 3. The dynamics (\ref{population}) of this population is described by the following equations
\begin{flalign}\label{eqn:per4_CRN2_3}
\begin{split}
\dot x_1 &= f_1 x_1 - p x_1 (r_1 + \beta r_2),\\
\dot x_2 &= f_2 x2 - p x_2 (r_2 + \beta r_3),\\
\dot x_3 &= f_3 x_3 - p x_3 (r_3 + \beta r_2),\\
\dot x_4 &= f_4 x_4 - p x_4 (r_4 + \beta r_3),\\
\dot r_1 &= c (\frac{x_1 r_1}{r_1 + \alpha r_2}) - b r_1,\\
\dot r_2 &= c (\frac{\alpha x_1 r_2}{r_1 + \alpha r_2} + \frac{x_2 r_2}{r_2 + \alpha r_3} + \frac{\alpha x_3 r_2}{r_3 + \alpha r_2} - b r_2,\\
\dot r_3 &= c (\frac{\alpha x_2 r_3}{r_2 + \alpha r_3} + \frac{x_3 r_3}{r_3 + \alpha r_2} +  + \frac{\alpha x_4 r_3}{r_4 + \alpha r_3}) - b r_3,\\
\dot r_4 &= c (\frac{x_4 r_4}{r_4 + \alpha r_3}) - b r_4.
\end{split}
\end{flalign}

\begin{figure}[ht]
    \centering
    \includegraphics[scale=0.4]{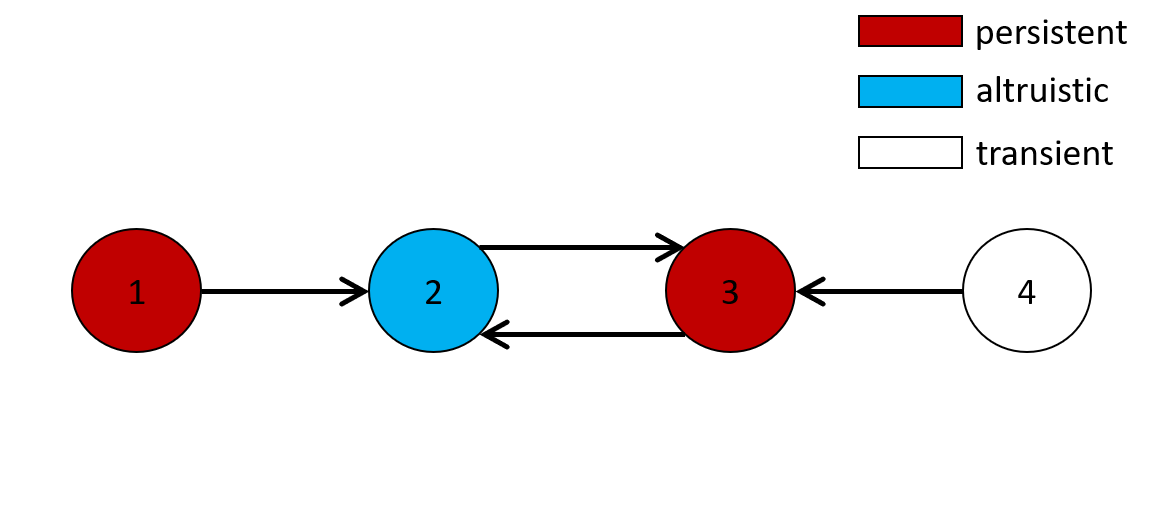}
    \caption{A fixed point when the newly emerging variant is connected to the variant 3 and becomes inactive}
    \label{fig:branch_cycle_4to3_inactive}
\end{figure}

The stable network obtained from this network is shown in figure \ref{fig:branch_cycle_4to3_inactive}.

The fixed points resulting in the network in figure \ref{fig:branch_cycle_4to3_inactive} are
\begin{flalign*}
    x_1 &= \frac{b f_1}{\beta c p}(1-\alpha), &x_2 &= 0, &x_3 &= \frac{b}{\beta c p}((\alpha - \beta)f_1 + \beta f_3),  &x_4 &= 0\\
    r_1 &= 0, &r_2 &= \frac{f_1}{\beta p}, &r_3 &= \frac{f_3 - f_1}{p}, &r_4 &= 0
\end{flalign*}

This network (\ref{fig:branch_cycle_4to3_inactive}) has the newly emerged variant being transient, and the resulting network is the same as the initial branch-cycle network.

\subsection{Computations for merging of two minimal networks}
Fig. \ref{fig:combined_networks} depicts three instances where two symmetric minimal networks are connected to each other to form three different types of networks.\\
The dynamics (\ref{population}) of this configuration is as follows:

\begin{flalign}\label{eqn:symm_joined}
\begin{split}
\dot x_1 &= f_1 x_1 - p x_1 (r_1 + \beta r_2 + \beta r_5),\\
\dot x_2 &= f_2 x2 - p x_2 r_2,\\
\dot x_3 &= f_3 x_3 - p x_3 (r_3 + \beta r_2),\\
\dot x_4 &= f_4 x_4 - p x_4 (r_4 + \beta r_2 + \beta r_5),\\
\dot x_5 &= f_5 x_5 - p x_5 r_5,\\
\dot x_6 &= f_6 x_6 - p x_6 (r_6 + \beta r_5),\\
\dot r_1 &= c (\frac{x_1 r_1}{r_1 + \alpha r_2 + \alpha r_5}) - b r_1,\\
\dot r_2 &= c (\frac{\alpha x_1 r_2}{r_1 + \alpha r_2 + \alpha r_5} + x_2 + \frac{\alpha x_3 r_2}{r_3 + \alpha r_2} + \frac{\alpha x_4 r_2}{r_4 + \alpha r_2 + \alpha r_5}) - b r_2,\\
\dot r_3 &= c (\frac{x_3 r_3}{r_3 + \alpha r_2}) - b r_3,\\
\dot r_4 &= c (\frac{x_4 r_4}{r_4 + \alpha r_2 + \alpha r_5}) - b r_4,\\
\dot r_5 &= c(\frac{\alpha x_1 r_5}{r_1 + \alpha r_2 + \alpha r_5} + \frac{\alpha x_4 r_5}{r_4 + \alpha r_2 + \alpha r_5} + x_5 + \frac{\alpha x_6 r_5}{r_6 + \alpha r_5}) - b r_5,\\
\dot r_6 &= c(\frac{x_6 r_6}{r_6 + \alpha r_5}) - b r_6.
\end{split}
\end{flalign}

The Jacobian of the system of equations \ref{eqn:symm_joined} at the fixed point shown in figure \ref{fig:combined_3} equals:
$$
J = 
\begin{pmatrix}
X & Y
\end{pmatrix}
$$
where
\begin{align*}
    X = \left(\begin{array}{ccccccc} f_{1}-f_{4} & 0 & 0 & 0 & 0 & 0 & 0\\ 0 & f_{2}-\frac{f_{4}}{\beta } & 0 & 0 & 0 & 0 & 0\\ 0 & 0 & 0 & 0 & 0 & 0 & 0\\ 0 & 0 & 0 & 0 & 0 & 0 & 0\\ 0 & 0 & 0 & 0 & f_{5} & 0 & 0\\ 0 & 0 & 0 & 0 & 0 & 0 & 0\\ 0 & 0 & 0 & 0 & 0 & 0 & -b\\ c & c & \frac{\alpha \,c\,f_{4}}{\alpha \,f_{4}+\beta \,f_{3}-\beta \,f_{4}} & c & 0 & 0 & 0\\ 0 & 0 & \frac{\beta \,c\,\left(f_{3}-f_{4}\right)}{\alpha \,f_{4}+\beta \,f_{3}-\beta \,f_{4}} & 0 & 0 & 0 & 0\\ 0 & 0 & 0 & 0 & 0 & 0 & 0\\ 0 & 0 & 0 & 0 & c & 0 & 0\\ 0 & 0 & 0 & 0 & 0 & c & 0 \end{array}\right),
\end{align*}
and
\begin{align*}
    Y = \scalemath{0.7}{\left(\begin{array}{ccccc} 0 & 0 & 0 & 0 & 0\\ 0 & 0 & 0 & 0 & 0\\ -\frac{b\,\left(\alpha \,f_{4}+\beta \,f_{3}-\beta \,f_{4}\right)}{c} & -\frac{b\,\left(\alpha \,f_{4}+\beta \,f_{3}-\beta \,f_{4}\right)}{\beta \,c} & 0 & 0 & 0\\ \frac{b\,f_{4}\,\left(\alpha -1\right)}{c} & 0 & \frac{b\,f_{4}\,\left(\alpha -1\right)}{\beta \,c} & \frac{b\,f_{4}\,\left(\alpha -1\right)}{c} & 0\\ 0 & 0 & 0 & 0 & 0\\ 0 & 0 & 0 & -\frac{b\,\beta \,f_{6}}{c} & -\frac{b\,f_{6}}{c}\\ 0 & 0 & 0 & 0 & 0\\ -\frac{b\,\left(\alpha \,f_{4}+\beta \,f_{3}-\beta \,f_{4}-\alpha \,\beta \,f_{3}+\alpha \,\beta \,f_{4}\right)}{\alpha \,f_{4}+\beta \,f_{3}-\beta \,f_{4}} & -\frac{\alpha \,b\,f_{4}}{\alpha \,f_{4}+\beta \,f_{3}-\beta \,f_{4}} & \frac{b\,\left(\alpha -1\right)}{\alpha } & b\,\left(\alpha -1\right) & 0\\ -\frac{\alpha \,b\,\beta \,\left(f_{3}-f_{4}\right)}{\alpha \,f_{4}+\beta \,f_{3}-\beta \,f_{4}} & -\frac{b\,\beta \,\left(f_{3}-f_{4}\right)}{\alpha \,f_{4}+\beta \,f_{3}-\beta \,f_{4}} & 0 & 0 & 0\\ 0 & 0 & -\frac{b\,\left(2\,\alpha -1\right)}{\alpha } & 0 & 0\\ 0 & 0 & 0 & 0 & 0\\ 0 & 0 & 0 & -\alpha \,b & -b \end{array}\right).}
\end{align*}

The characteristic polynomial of this Jacobian is
\begin{align*}
    \det(J - \lambda I) &= \frac{1}{c x_3} (\lambda^5 (\lambda + (f_4 - f_1))(\lambda^2 + b\lambda + cpx_6)[cx_3\lambda^4 + b(\alpha^2 br_2 + br_3\\
    &+ cx_3(1-\alpha))\lambda^3 + (\alpha bcf_4x_3 + \beta c^2 px_3x_4 + b^3r_3 + bcpx_3r_3)\lambda^2 \\
    &+ b^2cpr_3(\beta x_4 + x_3(1-\alpha))\lambda + b\beta c^2p^2x_3x_4]) 
\end{align*}
As all coefficients of this polynomial is positive, it cannot have real positive roots. Therefore this fixed point corresponds to a stable state of local immunodeficiency.\\

The Jacobian of the fixed point in figure \ref{fig:combined_1} is $J = (W X Y Z)$ where
\begin{align*}
    W = \left(\begin{array}{ccccc} f_{1}-f_{4} & 0 & 0 & 0 & 0\\ 0 & f_{2}-\frac{f_{4}-\beta \,f_{5}}{\beta } & 0 & 0 & 0\\ 0 & 0 & 0 & 0 & 0\\ 0 & 0 & 0 & 0 & 0\\ 0 & 0 & 0 & 0 & 0\\ 0 & 0 & 0 & 0 & 0\\ 0 & 0 & 0 & 0 & 0\\ \frac{c\,\left(f_{4}-\beta \,f_{5}\right)}{f_{4}} & c & \frac{\alpha \,c\,\left(f_{4}-\beta \,f_{5}\right)}{\alpha \,f_{4}+\beta \,f_{3}-\beta \,f_{4}+\beta ^2\,f_{5}-\alpha \,\beta \,f_{5}} & \frac{c\,\left(f_{4}-\beta \,f_{5}\right)}{f_{4}} & 0\\ 0 & 0 & \frac{\beta \,c\,\left(f_{3}-f_{4}+\beta \,f_{5}\right)}{\alpha \,f_{4}+\beta \,f_{3}-\beta \,f_{4}+\beta ^2\,f_{5}-\alpha \,\beta \,f_{5}} & 0 & 0\\ 0 & 0 & 0 & 0 & 0\\ \frac{\beta \,c\,f_{5}}{f_{4}} & 0 & 0 & \frac{\beta \,c\,f_{5}}{f_{4}} & c\\ 0 & 0 & 0 & 0 & 0 \end{array}\right),
\end{align*}
\begin{align*}
    X = \scalemath{0.7}{\left(\begin{array}{ccc} 0 & 0 & 0\\ 0 & 0 & 0\\ 0 & 0 & -\frac{b\,\left(\alpha \,f_{4}+\beta \,f_{3}-\beta \,f_{4}+\beta ^2\,f_{5}-\alpha \,\beta \,f_{5}\right)}{c}\\ 0 & 0 & \frac{b\,f_{4}\,\left(\alpha -1\right)}{c}\\ 0 & 0 & 0\\ 0 & 0 & 0\\ 0 & -b & 0\\ 0 & 0 & -\frac{b\,\left(f_{4}-\beta \,f_{5}\right)\,\left(\alpha \,f_{4}+\beta \,f_{3}-\beta \,f_{4}+\beta ^2\,f_{5}-\alpha \,\beta \,f_{3}+\alpha \,\beta \,f_{4}-\alpha \,\beta \,f_{5}-\alpha \,\beta ^2\,f_{5}+\alpha ^2\,\beta \,f_{5}\right)}{f_{4}\,\left(\alpha \,f_{4}+\beta \,f_{3}-\beta \,f_{4}+\beta ^2\,f_{5}-\alpha \,\beta \,f_{5}\right)}\\ 0 & 0 & -\frac{\alpha \,b\,\beta \,\left(f_{3}-f_{4}+\beta \,f_{5}\right)}{\alpha \,f_{4}+\beta \,f_{3}-\beta \,f_{4}+\beta ^2\,f_{5}-\alpha \,\beta \,f_{5}}\\ 0 & 0 & 0\\ \frac{\alpha \,c\,f_{5}}{f_{6}+\alpha \,f_{5}-\beta \,f_{5}} & 0 & \frac{b\,\beta \,f_{5}\,\left(\alpha -1\right)}{f_{4}}\\ \frac{c\,\left(f_{6}-\beta \,f_{5}\right)}{f_{6}+\alpha \,f_{5}-\beta \,f_{5}} & 0 & 0 \end{array}\right),}
\end{align*}
\begin{align*}
    Y = \left(\begin{array}{cc} 0 & 0\\ 0 & 0\\ -\frac{b\,\left(\alpha \,f_{4}+\beta \,f_{3}-\beta \,f_{4}+\beta ^2\,f_{5}-\alpha \,\beta \,f_{5}\right)}{\beta \,c} & 0\\ 0 & \frac{b\,f_{4}\,\left(\alpha -1\right)}{\beta \,c}\\ 0 & 0\\ 0 & 0\\ 0 & 0\\ -\frac{\alpha \,b\,\left(f_{4}-\beta \,f_{5}\right)}{\alpha \,f_{4}+\beta \,f_{3}-\beta \,f_{4}+\beta ^2\,f_{5}-\alpha \,\beta \,f_{5}} & \frac{b\,\left(\alpha -1\right)\,\left(f_{4}-\beta \,f_{5}\right)}{\alpha \,f_{4}}\\ -\frac{b\,\beta \,\left(f_{3}-f_{4}+\beta \,f_{5}\right)}{\alpha \,f_{4}+\beta \,f_{3}-\beta \,f_{4}+\beta ^2\,f_{5}-\alpha \,\beta \,f_{5}} & 0\\ 0 & -\frac{b\,\left(2\,\alpha -1\right)}{\alpha }\\ 0 & \frac{b\,\beta \,f_{5}\,\left(\alpha -1\right)}{\alpha \,f_{4}}\\ 0 & 0 \end{array}\right),
\end{align*}
and
\begin{align*}
    Z = \left(\begin{array}{cc} 0 & 0\\ 0 & 0\\ 0 & 0\\ \frac{b\,f_{4}\,\left(\alpha -1\right)}{c} & 0\\ 0 & 0\\ -\frac{b\,\beta \,\left(f_{6}+\alpha \,f_{5}-\beta \,f_{5}\right)}{c} & -\frac{b\,\left(f_{6}+\alpha \,f_{5}-\beta \,f_{5}\right)}{c}\\ 0 & 0\\ \frac{b\,\left(\alpha -1\right)\,\left(f_{4}-\beta \,f_{5}\right)}{f_{4}} & 0\\ 0 & 0\\ 0 & 0\\ -\frac{b\,f_{5}\,\left(\beta \,f_{6}+\alpha ^2\,f_{4}-\beta ^2\,f_{5}+\alpha \,\beta \,f_{5}-\alpha \,\beta \,f_{6}+\alpha \,\beta ^2\,f_{5}-\alpha ^2\,\beta \,f_{5}\right)}{f_{4}\,\left(f_{6}+\alpha \,f_{5}-\beta \,f_{5}\right)} & -\frac{\alpha \,b\,f_{5}}{f_{6}+\alpha \,f_{5}-\beta \,f_{5}}\\ -\frac{\alpha \,b\,\left(f_{6}-\beta \,f_{5}\right)}{f_{6}+\alpha \,f_{5}-\beta \,f_{5}} & -\frac{b\,\left(f_{6}-\beta \,f_{5}\right)}{f_{6}+\alpha \,f_{5}-\beta \,f_{5}} \end{array}\right).
\end{align*}
We will present now an exact example with the stable state of local immunodeficiency. Let the system’s parameters have the following values $f_1 = 0.25, f_2 = 0.3, f_3 = 0.35, f_4 = 0.3, f_5 = 0.35, f_6 = 0.4, c = 1, p = 2, \alpha = 1/3, \beta = 1/9, b = 3$. One can compute the corresponding Jacobian numerically and confirm that all the eigenvalues are either real negative or complex with negative real parts. It follows by continuity that there exists a positive measure set in the parameter space such that for any point (a set of parameters) the corresponding state of local immunodeficiency is stable.\\

Finally, the Jacobian computed at the fixed point shown in figure \ref{fig:combined_2} is $J = X Y$ where
\begin{align*}
    X = \left(\begin{array}{cccccc} f_{1}-f_{6}-\beta \,f_{2} & 0 & 0 & 0 & 0 & 0\\ 0 & 0 & 0 & 0 & 0 & 0\\ 0 & 0 & 0 & 0 & 0 & 0\\ 0 & 0 & 0 & f_{4}-f_{6}-\beta \,f_{2} & 0 & 0\\ 0 & 0 & 0 & 0 & f_{5}-\frac{f_{6}}{\beta } & 0\\ 0 & 0 & 0 & 0 & 0 & 0\\ 0 & 0 & 0 & 0 & 0 & 0\\ \frac{\beta \,c\,f_{2}}{f_{6}+\beta \,f_{2}} & c & \frac{\alpha \,c\,f_{2}}{f_{3}+\alpha \,f_{2}-\beta \,f_{2}} & \frac{\beta \,c\,f_{2}}{f_{6}+\beta \,f_{2}} & 0 & 0\\ 0 & 0 & \frac{c\,\left(f_{3}-\beta \,f_{2}\right)}{f_{3}+\alpha \,f_{2}-\beta \,f_{2}} & 0 & 0 & 0\\ 0 & 0 & 0 & 0 & 0 & 0\\ \frac{c\,f_{6}}{f_{6}+\beta \,f_{2}} & 0 & 0 & \frac{c\,f_{6}}{f_{6}+\beta \,f_{2}} & c & c\\ 0 & 0 & 0 & 0 & 0 & 0 \end{array}\right),
\end{align*}
\begin{align*}
    Y = \left(\begin{array}{cccccc} 0 & 0 & 0 & 0 & 0 & 0\\ 0 & \frac{b\,f_{2}\,\left(\alpha -1\right)}{c} & 0 & 0 & 0 & 0\\ 0 & -\frac{b\,\beta \,\left(f_{3}+\alpha \,f_{2}-\beta \,f_{2}\right)}{c} & -\frac{b\,\left(f_{3}+\alpha \,f_{2}-\beta \,f_{2}\right)}{c} & 0 & 0 & 0\\ 0 & 0 & 0 & 0 & 0 & 0\\ 0 & 0 & 0 & 0 & 0 & 0\\ 0 & 0 & 0 & 0 & -\frac{b\,f_{6}}{c} & -\frac{b\,f_{6}}{\beta \,c}\\ -b & 0 & 0 & 0 & 0 & 0\\ 0 & -\frac{b\,\left(f_{3}+\alpha \,f_{2}-\alpha \,f_{3}-\beta \,f_{2}+\alpha \,\beta \,f_{2}\right)}{f_{3}+\alpha \,f_{2}-\beta \,f_{2}} & -\frac{\alpha \,b\,f_{2}}{f_{3}+\alpha \,f_{2}-\beta \,f_{2}} & 0 & 0 & 0\\ 0 & -\frac{\alpha \,b\,\left(f_{3}-\beta \,f_{2}\right)}{f_{3}+\alpha \,f_{2}-\beta \,f_{2}} & -\frac{b\,\left(f_{3}-\beta \,f_{2}\right)}{f_{3}+\alpha \,f_{2}-\beta \,f_{2}} & 0 & 0 & 0\\ 0 & 0 & 0 & -b & 0 & 0\\ 0 & 0 & 0 & 0 & -b & -\frac{b}{\alpha }\\ 0 & 0 & 0 & 0 & 0 & -\frac{b\,\left(\alpha -1\right)}{\alpha } \end{array}\right),
\end{align*}
In this case also, we present here an exact numerical example with a stable state of local immunodeficiency. Let the system’s parameters assume the following values $f_1 = 0.25, f_2 = 0.3, f_3 = 0.35, f_4 = 0.3, f_5 = 0.35, f_6 = 0.4, c = 1, p = 2, \alpha = 1/3, \beta = 1/9, b = 3$. One can compute the corresponding Jacobian numerically and confirm that all the eigenvalues are either real negative or complex with negative real parts. Once again, it follows by continuity that there exists a positive measure set in the parameter space where the state of a local immunodeficiency is stable.\\

 \bibliographystyle{elsarticle-num} 
 \bibliography{sn-bibliography, grants1}





\end{document}